\documentclass[prd,showpacs,nofootinbib,preprintnumbers,preprint]{revtex4}
\usepackage{amsmath}  \usepackage{amsfonts}
\usepackage{graphicx,
xcolor
}
\renewcommand{\theequation}{\arabic{section}.\arabic{equation}}
\def\be{\begin{equation}}
\def\ee{\end{equation}}
\def\ba{\begin{eqnarray}}
\def\ea{\end{eqnarray}}
\def\nn{\nonumber}
\def\lb{\label}
\def\dfrac{\displaystyle\frac}
\def\bb{\bibitem}
\def\v{\overline{v}}
\def\E{{\cal E}}
\def\p{\hat\varphi}
\def\x{\overline{x}}
\def\Pa{\overline{\Pi}_1}
\def\Pb{\overline{\Pi}_2}

\begin{document}
%\begin{titlepage}
%\date{}
\title{\begin{flushright}\begin{small}    LAPTH-026/18
\end{small} \end{flushright} \vspace{1.5cm}
Stationary double black hole\\ without naked ring singularity}
\author{G\'erard Cl\'ement} \email{gerard.clement@lapth.cnrs.fr}
\affiliation{LAPTh, Universit\'e Savoie Mont Blanc, CNRS, 9 chemin de Bellevue, \\
BP 110, F-74941 Annecy-le-Vieux cedex, France}
\author{Dmitri Gal'tsov} \email{galtsov@phys.msu.ru}
\affiliation{Faculty of Physics,
Moscow State University, 119899, Moscow, Russia,\\Kazan Federal University, 420008 Kazan, Russia}

%\author{G\'erard Cl\'ement$^a$\thanks{Email: gclement@lapth.cnrs.fr},
%Dmitri Gal'tsov$^{b,c}$\thanks{Email: galtsov@phys.msu.ru} \\ \\
%$^a$ {\small LAPTh, Universit\'e Savoie Mont Blanc, CNRS, 9 chemin de Bellevue,} \\
%{\small BP 110, F-74941 Annecy-le-Vieux cedex, France} \\
%$^b$ {\small Department of Theoretical Physics, Faculty of Physics,}\\
%{\small Moscow State University, 119899, Moscow, Russia }\\
%$^c$ {\small  Kazan Federal University, 420008 Kazan, Russia}}

\begin{abstract}
Recently double black hole vacuum and electrovacuum metrics attracted attention as
exact solutions suitable for visualization of ultra-compact objects beyond the Kerr paradigm.
However, many of the proposed systems are plagued with ring curvature singularities.
Here we present a new simple solution of this type which is asymptotically Kerr, has zero electric
and magnetic charges, but is endowed with magnetic dipole moment and electric quadrupole moment.
It is manifestly free of ring singularities, and contains only a mild string-like singularity
on the axis corresponding to a distributional energy-momentum tensor.
Its main constituents are two extreme co-rotating black holes carrying equal electric and
opposite magnetic and NUT charges.
\end{abstract}
%\end{titlepage}
\maketitle

\setcounter{page}{2}

\setcounter{equation}{0}
\section{Introduction}
Binary black holes  became an especially hot topic after the discovery
of the first gravitational wave signal from merging black holes \cite{Barack:2018yly}. A particular
interest lies in determining their observational features other than emission
of strong gravitational waves. Such features include gravitational lensing
and shadows which presumably can be observed in experiments such as
the Event Horizon Telescope and future space projects. For these experiments to
have sense, one needs to model plausible alternatives to the Kerr paradigm \cite{Yagi}.
Most proposed scenarios use phenomenologically constructed metrics such as deformed Kerr, or ``bumpy''
black holes, not necessarily satisfying the Einstein equations, or exact solutions for black holes
with hair, and black holes in modified gravity to study lensing/shadow pictures  \cite{Cunha:2018acu}.

Some popular lensing/shadow models  use the double Kerr solution of Kramer and Neugebauer
\cite{Kramer1980, Costa:2010zzg}  describing two rotating Kerr black holes sitting on the same axis,
as well as more general vacuum and electrovacuum solutions for double black holes (DBH)
\cite{Cunha:2018gql,Cunha:2018cof} . One should be warned, however, that typically DBH solutions contain strong
curvature singularities in the physical region, which are often overlooked or ignored \cite{Johannsen:2013rqa}.
Solutions with strong naked curvature singularities should certainly be rejected as unphysical.
At the same time, in view of uniqueness theorems for Kerr black hole, to go beyond the Kerr paradigm one
{\em should} tolerate violation of some standard assumptions, preferably those which were not
rigorously proven. In our opinion, mild violations of cosmic censorship, rejecting strong naked curvature
singularities, while admitting distributional ones, such as conical singularities associated with
infinitely thin cosmic strings, deserve to be explored more closely. From the theory of cosmic strings.
it is known that such singularities can be removed by the introduction of suitable additional matter,
so their presence in DBHs can be regarded just as evidence for extra matter, maybe exotic. In view of
unsolved dark matter and dark energy problems, such metrics should not be rejected.

With this motivation, we present here a detailed discussion of a new DBH solution of Einstein-Maxwell
equations (a short presentation was given in \cite{Clement:2017kas}) devoid of strong curvature singularities
but containing a string singularity between the constituent black holes. This solution was first constructed,
not by the soliton technique which is now the main tool in the domain of exact solutions, but by applying
an original approach of one of the authors \cite{GC98} which opens a way to generate rotating electrovacuum
solutions generically not accessible by soliton dressing.

An early solution, the magnetic dipole of Bonnor (1966) \cite{Bonnor:1966}, was later reinterpreted as a
{\em dihole}: a static system of two extremal black holes with opposite magnetic charges
\cite{Emparan:1999au,Emparan:2001bb}. Another early family of stationary vacuum solutions was that of
``deformed'' Kerr metrics given by Tomimatsu and Sato \cite{Tomimatsu:1972zz} (TS)
with the integer deformation parameter
$\delta$, such that the Kerr metric is reproduced for $\delta=1$. Its $\delta=2$ member was later
interpreted as a DBH. In fact, the static version of the Tomimatsu-Sato solution with $\delta=2$ (TS2) was
known since the papers by Bach and Weyl (1922) \cite{BW} , Darmois (1927) \cite{darmois} , Zipoy (1966)
\cite{Zipoy1966} and Voorhees (1971) \cite{Voorhees:1971wh} (ZV). These solutions, expressed in
prolate spheroidal coordinates ($x,y$), exhibit directional singularities at $x=1, y=\pm 1$ \cite{gibbons73}.
The  two-surface  nature of the ``points'' $x=1, y=\pm 1$ of the TS2 solution, hinted at in
\cite{Ernst,Economou}, was then explicitly demonstrated by Papadopoulos et al. in 1981
\cite{Papadopoulos:1981wr} showing that these are black hole horizons. Still, misinterpretation persisted
in some papers (see e.g. \cite{Manko:1999xg}), until it was unambiguosly proven by Kodama and Hikida
\cite{Kodama:2003ch} that both  ZV2 and TS2 are DBH solutions indeed, though singular 
(for modern analysis and applications also see \cite{Gegenberg:2010np}). The Tomimatsu-Sato family,
apart from the ``deformation'' parameter $\delta$, have only two physical parameters which can be chosen
as the total mass and angular momentum, as for the Kerr metric. The Kerr uniqueness is circumvented in 
this case because the TS2 metric contains a naked ring singularity in the domain of outer communications
\cite{Tomimatsu:1972zz}. Our solution can be interpreted as a new rotating electrovacuum
extension of the ZV2 solution different from TS2 as well as from previously known electrovacuum extensions
of TS2 \cite{Ernst1,Manko:1999xg}, which are also plagued with strong naked singularities.

A new hunt for double-center solutions started with Kramer and Neugebauer \cite{Kramer1980}, who applied
B\"acklund transformations  \cite{Hoenselaers:1985qk} to construct
two-black hole solutions with the individual masses, angular momenta, and NUT charges of the constituents,
together with the distance between them, as free parameters. Soon after it was realized that the
inverse scattering method (ISM) of Belinski and Zakharov \cite{Belinsky:1979mh, Belinski:2001ph} is a
more universal and convenient tool to produce DBH solutions via soliton dressing of black-hole backgrounds
(for a review and further references see an accessible introduction by Herdeiro et al. \cite{Herdeiro:2008kq},
an extensive book by Griffiths and Podolsky \cite{GP}, and a more recent paper by Alekseev
\cite{Alekseev:2017zuh}). It is worth noting that the method used in the current paper, though it
generates only a one-parameter family of new solutions, is applicable in the case of generic non-analytic
metrics where the ISM does not work. In the case of applicability of ISM both methods lead to identical results.

One of the main problems in the DBH theory was the search for equilibrium configurations of two vacuum or
charged black holes. It has been known for some time that equilibrium of two centers generically carrying
masses and electric (magnetic) charges is possible only in the case when masses and charges satisfy the
so-called no-force conditions \cite{Breitenlohner:1987dg}. The corresponding solutions are characterized by 
conformally flat three-metrics, so that consequently multi-center solutions may exist for any positions of 
the centers  (Majumdar-Papapetrou multi-black hole solution). But, two-center solutions
can also exist for a certain separation between the centers, with a quite different relationship between
the parameters \cite{Bonnor:1993,Perry}, though these have naked singularities.

An intriguing question was whether the gravitational spin-spin interaction, which is repulsive for parallel
spins, can overcome gravitational attraction. The extremal co-rotating Kerr black holes seem to be most
favored for this. General conditions for force balance were formulated by Tomimatsu and collaborators
\cite{Tomimatsu:1981bc,Kihara:1981mx,Tomimatsu:1983qc}, but the first found rotating configurations obeying
these conditions were shown to contain strong naked singularities. A thorough investigation of double-Kerr
solutions by Dietz and Hoenselaers \cite{Dietz1985} led to the conclusion that the balance is not possible
for two black holes with regular horizons, unless hyperextreme objects (which are not black holes, but
naked singularities) are involved. Later, several non-existence theorems for stationary balanced vacuum
two-black hole systems (including extremal black holes) were formulated by Neugebauer and Hennig
\cite{Neugebauer:2009su,Hennig:2011fp,Neugebauer:2011qb}, see also \cite{Chrusciel:2011iv}.
The conclusion following from this analysis is that to balance two asymptotically flat black holes one
needs conical singularities (a cosmic string) on the axis between the centers. Similar conclusions
were derived in the electrovacuum case \cite{Alekseev:2012au}.

An additional restriction comes from the requirement that the string be non-rotating, since otherwise
it will be surrounded by a region where $g_{\varphi\varphi}$ changes sign, implying the existence of
closed timelike curves (CTC). For two rotating black holes this causes additional restrictions on the
parameters, excluding in particular the possibility of NUT charges of the constituents (``axis conditions'').
Here we do not introduce these restrictions, and allow for the possibility of rotating cosmic strings.
In so doing, we enlarge the family of metrics with a distributional Ricci tensor, which indicates that
new forms of matter should be involved. On the other hand,
we exclude ring curvature singularities in the physical region, since these can in no case be smoothed
out by additional matter sources.

\setcounter{equation}{0}
\section{Constructing the solution}
\subsection{The seed}
Our new solution was constructed using the original generating technique of \cite{GC98},
with a seed belonging to the Weyl stationary axisymmetric class, the static
ZV2 vacuum solution. In the standard Weyl-Papapetrou parametrization
 \be\lb{weyl}
ds^2 = -F(dt-\omega d\varphi)^2 +
F^{-1}[e^{2k}(d\rho^2+dz^2)+\rho^2d\varphi^2],
 \ee
the ZV2 solution \cite{Zipoy1966,Voorhees:1971wh}, first given by Darmois \cite{darmois}, is
characterized by
 \be
F=\left(\dfrac{x-1}{x+1}\right)^2, \quad e^{2k} =
\left(\dfrac{x^2-1}{x^2-y^2}\right)^4, \quad \omega = 0,
 \ee
where the prolate spheroidal coordinates $(x,y)$ are related to the
Weyl coordinates by
 \be\lb{sphero}
\rho=\kappa(x^2-1)^{1/2}(1-y^2)^{1/2}, \quad z=\kappa xy
 \ee
(the positive constant $\kappa$ setting the length scale).

This solution is asymptotically (as $x\to \infty$) flat and has the Schwarzschild
mass $M=2\kappa$, which can be easily seen by using the asymptotic identification
 \be \lb{asx} x\approx r/\kappa - 1 + {\rm O}(1/r) ,\quad   y=\cos\theta.
 \ee
To reveal the singularities, one computes the Kretschmann scalar
$K=R_{\mu\nu\lambda\tau}R^{\mu\nu\lambda\tau}$, obtaining
 \be\lb{kret}
K= \frac{192(x^2-y^2)^5}{\kappa^2(x^2-1)^6(x+1)^8}\left[x^4-4 x^3+(7-4 y^2)x^2+(10y^2-6)x-7y^2+3\right].
 \ee
The coordinate region $x\geq 1,\, |y|\leq 1$ thus presents an asymptotically flat
space-time with a curvature singularity at $x=1$ and $|y|<1$, which corresponds
to an open set in the Weyl coordinates $\rho=0,\,-\kappa<z<\kappa$ (the singular rod).

The boundary points of the rod $z=\pm \kappa$ are directional singularities.
If one approaches  $z=\kappa,\,\rho=0$ sending $x\to 1, y\to 1$ and keeping the
ratio $X^2=(1-y^2)/(x^2-1)$ fixed (with  $Y=y/x\to 1$ ), one finds that the Kretschmann
scalar depends on $X$,
 \be\lb{kretH}
K_H=\lim_{Y\to\pm 1}K=\frac{3(X^2+1)^6}{4\kappa^4},
\ee
and diverges as $X\to\infty$. The same value of $K$ will be obtained
approaching $z=-\kappa$ with fixed $X$, in which case $Y\to -1$. Following \cite{Kodama:2003ch}
and using $X,\,Y$ as new coordinates,
 \be\lb{XY}
X=\sqrt{\frac{1-y^2}{x^2-1}}, \;\; Y = \frac{y}x, \quad x =
\sqrt{\frac{X^2+1}{X^2+Y^2}}, \;\; y =
Y\,\sqrt{\frac{X^2+1}{X^2+Y^2}} ,
 \ee
one can see that the infinite value $X=\infty$ corresponds to the singular
open set $\rho=0,\, -\kappa< z< \kappa$, while the limiting ``points'' $\rho=0,\, z= \pm \kappa$
are actually two-surfaces $Y = \pm 1$. Using the relations
 \ba\lb{XY1}
&& x^2-1 = \frac{1-Y^2}{X^2+Y^2}, \quad 1-y^2 =
\frac{X^2(1-Y^2)}{X^2+Y^2}, \quad x^2-y^2 =
\frac{(X^2+1)(1-Y^2)}{X^2+Y^2}, \nn\\
&& \rho = \frac{\kappa X(1-Y^2)}{X^2+Y^2},
\quad z = \frac{\kappa Y(X^2+1)}{X^2+Y^2}, \\
&& d\rho^2+dz^2 = \kappa^2(x^2-y^2)\left(\frac{dx^2}{x^2-1} +
\frac{dy^2}{1-y^2}\right) =
\kappa^2(x^2-y^2)^2\left[\frac{dX^2}{(X^2+1)^2} +
\frac{dY^2}{(1-Y^2)^2}\right]
,\nn
 \ea
one can rewrite the metric in the vicinity of $Y= \pm 1$ as
 \be\lb{ZVH}
ds_{\pm}^2 = \frac1{(X^2+1)^2}\left[-\frac{(1 \mp Y)^2}{4}\,dt^2 +
\frac{4\kappa^2}{(1\mp Y)^2} \,dY^2\right] +
16\kappa^2\left[\frac{dX^2}{(X^2+1)^4}+X^2 d\varphi^2\right]\,.
 \ee
Apparently, these metrics describe two extremal black holes with
degenerate horizons $H_{\pm}$ at $Y=\pm 1$, and finite horizon area
\cite {Kodama:2003ch}
 \be
{\cal A}_H = 32\pi\kappa^2\int_0^\infty \frac{XdX}{(X^2+1)^2} =
16\pi\kappa^2.
 \ee
Kodama and Hikida \cite{Kodama:2003ch} have constructed an
analytic continuation through the horizons to other Lorentzian
sectors $|y|>1, |x|<1$. However these horizons are not regular, but share
the ring-like curvature singularity $X\to\infty$ as common boundary. They
have also shown that the Komar mass of this singularity
is equal to the total mass $M=2\kappa$ (which is unusual for
naked singularities typically corresponding to negative mass), so that
the black holes themselves are massless.

The ZV2 metric has two Killing symmetries ($\partial_t$ and $\partial_\varphi$)
and no second-order Killing tensors, so the geodesic equations and the wave
equations are not separable. While equatorial motion can be explored in a
closed form, the non-equatorial orbits can be studied only numerically and
generically exhibit chaotic features.  Recently this metric attracted
attention as an alternative to standard black holes and the corresponding
shadows were constructed \cite{Cunha:2018acu}.

Rotating vacuum generalizations were constructed for integer $\delta$ by
Tomimatsu and Sato \cite{Tomimatsu:1972zz}, who showed that the $\delta=2$ rotating
solution (TS2) has a naked ring singularity. As discussed
by Gibbons and Russel-Clark \cite{gibbons73}, it also has a causal
boundary ($g_{\varphi\varphi}=0$), and a non-curvature Misner-string
singularity at $x=1$. The subsequent analysis of Kodama and Hikida
\cite{Kodama:2003ch} revealed that the segment $x=1$ is generically a line of
conical singularities (cosmic string) connecting two degenerate, topologically
spherical horizons at $x=\pm y=1$. Thus, the rotating TS2 solution is
more regular than the ZV2 solution.

\subsection{Cl\'ement transformation}
The four-dimensional stationary Einstein-Maxwell equations are invariant under
an $SU(2,1)$ group of transformations \cite{Kinnersley:1978pz,Kinnersley:1978}. These transformations map
asymptotically flat monopole solutions into monopole solutions, and so
cannot be used to transform an axisymmetric static monopole solution into a
rotating monopole--dipole solution. It was shown in \cite{GC98} that
this goal could be achieved by combining $SU(2,1)$ transformations
changing the asymptotic behavior with linear coordinate transformations
in the plane of the two Killing vectors, leading to a special finite Geroch
transformation.

More precisely, the rotation--generating transformation is the product
\begin{equation}\label{sig}
\Sigma = \Pi^{-1}\,{\cal R}(\Omega)\,\Pi
\end{equation}
of three successive transformations, two ``vertical'' transformations
$\Pi\,,\Pi^{-1} \in$ SU(2,1) acting on the space of the complex Ernst potentials
${\cal E}$ and $\psi$, and a ``horizontal'' global coordinate transformation
${\cal} R(\Omega)$ acting on the Killing 2--plane. The transformation $\Pi:\;
({\cal E}, \psi) \leftrightarrow (\hat{\cal E},\hat{\psi})$ with
\begin{equation}\label{inv}
\hat{\cal E} = \frac{-1 + {\cal E} \pm 2 \psi}{1 - {\cal E} +\pm2 \psi}\,, \quad
\hat{\psi} = \mp\frac{1 + {\cal E}}{1 - {\cal E} \pm 2 \psi}\,.
\end{equation}
leads from an asymptotically flat monopole seed solution to one which asymptotes
to $AdS_2 \times S^2$, i.e, is asymptotically Bertotti-Robinson (BR)-like.
The global coordinate transformation ${\cal R}(\Omega,\gamma(\Omega))$ is the
product of the transformation to a uniformly rotating frame and of a time
dilation,
\begin{equation}\label{R}
d\varphi = d\varphi' + \Omega\gamma\,dt'\,, \qquad dt = \gamma\,dt'\,.
\end{equation}
This does not modify the leading asymptotic behavior of asymptotically
BR--like metrics, so that the last transformation $\Pi^{-1}$ in (\ref{sig})
then leads to a new asymptotically flat solution with a dipole gravimagnetic
moment proportional to $\Omega$, ie. a rotating solution. In the case of
a vacuum seed solution, the parameter $\gamma(\Omega)$ can be
chosen so that this new solution has no monopole electromagnetic charges,
but it will generically (except in the case of the Schwarzschild seed, which
leads to the neutral Kerr solution) have a dipole magnetic moment and a
quadrupole electric moment.

\subsection{New solution}
The new rotating solution generated from the ZV2 solution by the
transformation $\Sigma$ can be given in terms of the Kinnersley
potentials\footnote{These differ from those given in (37) of
\cite{GC98} by the $\varepsilon$, related to the charge conjugation
$\pm$ in (\ref{inv}), by a change of the sign of $q$, and by a
common rescaling by a function of $x$.}:
 \ba\lb{kin}
U &=& p\dfrac{x^2+1}{2x} + iqy, \qquad V=\varepsilon(W-1),\nn\\
W &=& 1 + \dfrac{q^2}2\dfrac{1-y^2}{x^2-1} - i\dfrac{pq}2\dfrac{y}x,
\qquad (\varepsilon^2=1),
 \ea
related to the complex Ernst potentials by
 \be
{\cal E} = (U-W)/(U+W), \qquad \psi = V/(U+W).
 \ee
The real parameters $p$ and $q$ are related by $p=\sqrt{1-q^2}$ so
that, just as the TS2 solution, this family of solutions depends on
the single dimensionless rotation parameter $q$ (proportional to $\Omega$).
The potentials of the ZV2 solution are recovered for $q=0$.

The form (\ref{kin}) of the solution is only implicit. Dualization
of the imaginary part of the scalar Ernst potentials to vector
potentials leads to the explicit metric
 \ba
ds^2 &=& -
\frac{f}\Sigma\left(dt-\frac{\kappa\Pi}{f}\,d\varphi\right)^2 +
\kappa^2\Sigma\left[e^{2\nu}\left(\frac{dx^2}{x^2-1}
+ \frac{dy^2}{1-y^2}\right)\right. \nn\\
&& \left. \quad + f^{-1}(x^2-1)(1-y^2)d\varphi^2\right], \lb{anmet}
\ea
where $\varphi$ is periodic with period $2\pi$ as before, and
the Weyl metric functions are split as follows:
 \be
F = \frac{f}\Sigma, \quad \omega = \frac{\kappa\Pi}{f}, \quad e^{2k}
= \frac{fe^{2\nu}}{x^2-y^2},
 \ee
with
 \ba
f &=& \frac{p^2(x^2-1)^2}{4x^2} - \frac{q^2x^2(1-y^2)}{x^2-1}, \lb{f}\\
\Sigma &=& \left[\frac{px^2+2x+p}{2x} +
\frac{q^2(1-y^2)}{2(x^2-1)}\right]^2 +
q^2\left(1-\frac{p}{2x}\right)^2y^2, \lb{Sig}\\
e^{2\nu} &=& \frac{4x^2(x^2-1)^2}{p^2(x^2-y^2)^3}, \lb{e2nu}
\ea
Note that $\Sigma$ is positive definite. The rotation function $\omega$ is proportional to the
second order polynomial in $(1-y^2)$:
\be \Pi = \Pi_1(x)(1-y^2) + \Pi_2(x)(1-y^2)^2,\lb{Pi}
\ee
with $x$-dependent coefficients
\ba
\Pi_1 &=& -\frac{q}{2p}\left\{\frac{(px+2)[4x^2+p^2(x^2-1)]}{x^2} +
\frac{4p(1+p^2)x+8 + p^2  - p^4}{x^2-1}\right\}, \lb{Pi1}\\
\Pi_2 &=& -\frac{q^3}2\left[\frac{p}{4x^2} +
\frac{2x-p}{x^2-1}\right]. \lb{Pi2}
\ea
This is a stationary axisymmetric solution of the Einstein-Maxwell equations,
whose electromagnetic part is given by the four-potential
$A =A_0(x,y) dt+A_\varphi(x,y) d\varphi$, with
\be A_0 =\frac{\varepsilon \v}{\Sigma},\quad A_\varphi =\frac{\varepsilon\kappa\Theta}{\Sigma},\lb{anem}\ee
where
\be\v = \frac{q^2}4\left\{-\frac{p(2x-p)}{x^2} +
\left[\frac{p(2x-p)}{x^2} + \frac{px^2+2x+p}{x(x^2-1)}\right](1-y^2)
+ \frac{q^2(1-y^2)^2}{(x^2-1)^2} \right\}, \lb{Voorhees:1971wh}
\ee
$\varepsilon=\pm 1$, and $\Theta$ is the third order polynomial in $1-y^2$:
\be
\Theta = \Theta_1(x)(1-y^2) + \Theta_2(x)(1-y^2)^2 + \Theta_3(x)(1-y^2)^3,\lb{Theta}
\ee
with $x$-dependent coefficients
\begin{align}
\Theta_1 =& -\frac{q}{4p} \left\{\frac{p\left[p^2x^3 + 5px^2 -
(8-4p^2+p^4)x + 2p-3p^3\right]}{x^2}
+ \frac{(16-p^2+p^4)px + 8+7p^2+p^4}{x^2-1}\right\},\lb{Theta1} \\
\Theta_2 =&
-\frac{q^3}{8p}\left[\frac{p(4x^3-3px^2+2p^2x+5p)}{x^2(x^2-1)}
+ \frac{2(4x+3p+p^3)}{x(x^2-1)^2}\right],\lb{Theta2} \\
\Theta_3 =& -\frac{q^5}{8x(x^2-1)^2}. \lb{Theta3}
 \end{align}

This solution is different from the TS2 vacuum solution, but turns
out to coincide with a subclass of the four-parameter family of
solutions of the Einstein-Maxwell equations constructed by Manko et
al. in \cite{manko00}, corresponding to the two constraints on
their parameters (here indexed with $M$):
 \be
\delta_M = 0 \quad (\mu_M = -m_Mb_M,\;d_M=k_M^2), \quad (a_M-b_M)d_M = m_M^2b_M.
 \ee
The non-vanishing Manko et al. parameters are related to ours by
 \be
k_M = \kappa, \quad m_M = \frac{2\kappa}p, \quad b_M =
-\frac{\kappa pq}2, \quad a_M = -\frac{2\kappa q}p(1+p^2/4),
 \ee
and $m_Mb_M=-\kappa^2q$, $a_M-b_M=-2\kappa q/p$. The correspondence
between their metric functions (23) and ours is
 \be
\frac{f}{E_M} = \frac{\Sigma}{D_M} =
-\kappa\frac{\Pi_1+(1-y^2)\Pi_2}{F_M} =
\frac{p^2}{64\kappa^8x^2(x^2-1)^2}.
 \ee

%%%%%%%%%%%%%%%%%%%%%%%%%%%%%%%%%%%%%%5
\setcounter{equation}{0}
\section{Physical properties}
\subsection{Asymptotics}
The metric (\ref{anmet}) is
asymptotically (for $x\to\infty$) Minkowskian, which can be seen by introducing spherical
coordinates (\ref{asx}):
\begin{align}
F &\sim 1-\frac{4\kappa}{pr}+O\left(\frac1{r^2}\right),\\
\omega &\sim -\frac{2 q \kappa^2 (p^2+4)\sin^2\theta} {p^2r}+O\left(\frac1{r^2}\right),\\
g_{rr}&\sim 1+\frac{4\kappa}{pr}+O\left(\frac1{r^2}\right),\\
g_{\theta\theta}&\sim r^2,\quad g_{\varphi\varphi}\sim r^2\sin^2\theta .
\end{align}
The associated mass $M$ and angular momentum $J$ are
 \be\lb{asMJ}
M = \frac{2\kappa}p, \quad J = \frac{\kappa^2 q(4+p^2)}{p^2}.
 \ee
The electromagnetic potential exhibits the following asymptotic behavior:
\begin{align}
A_t &\sim \frac{\varepsilon\kappa^3 q^2(1-3\cos^3\theta)}{pr^2} +O\left(\frac1{r^3}\right),\\
A_\varphi &\sim -
\frac{\varepsilon\kappa^2 q\sin^2\theta}{r} +O\left(\frac1{r^2}\right).
\end{align}
It follows that the total electric and magnetic charges are zero, while there are a magnetic dipole
moment $\mu$, and a quadrupole electric moment $Q_2$:
\be
   \mu = \varepsilon\kappa^2q, \qquad Q_2 = -\varepsilon\kappa^3q^2/p.
\ee
The ratio $|\mu/J|$ is bounded above by $1/5$, in agreement with the
Barrow-Gibbons bound \cite{BG} $|\mu/J|\le1$, while the ratio $|J|/M^2$
satisfies the Kerr-like bound
 \be\lb{bound}
|J|/M^2 = |q|(1+p^2/4) \le 1.
 \ee
The upper bound in (\ref{bound}) is attained in the limit $p\to0$
with $M$ fixed, meaning also $\kappa = pM/2 \to 0$. Accordingly one
must first, as in the case of the Kerr or Kerr-Newman solutions,
rescale the radial coordinate $x$ by
 \be\lb{barx}
x = 2\frac{\x}p
 \ee
before taking the
limit $p\to0$, which for fixed $\x$ sends $x$ to infinity. It
follows that the Kinnersley potentials (\ref{kin}) go over to
 \be
U_{\rm lim} = \x \pm iy, \quad V_{\rm lim} = 0, \quad W_{\rm lim} =
1,
 \ee
which are those of the extreme Kerr metric. So the rescaled solution interpolates between two
limiting vacuum solutions, ZV2 for $q\to 0$ and extreme Kerr for $q\to 1$.

We will show later that, like ZV2, our solution, initially defined in
the domain $x\in]1,\infty), y\in[-1,1]$, can be extended through the
horizons at $x=|y|=1$ beyond this region. But let us first explore the solution
with decreasing $x$ step by step, starting from the region of large $x$ and
$y\in[-1,1]$ where $f>0$, $g_{\varphi\varphi}>0$.

\subsection{Absence of ring singularity}

The first obvious singularities appearing in rotating solutions
generated by the method proposed in \cite{GC98} are Kerr-like
ring singularities corresponding to zeroes of
$\Sigma(x,y)$ ($\Sigma$ is the sum of two squares, so
$\Sigma(x,y)=0$ actually corresponds to two equations).
To the difference of the TS2 vacuum solution, the present solution
is free from a naked ring singularity, as it is clear
from (\ref{Sig}) that $\Sigma$ admits the lower bound
 \be
\Sigma(x,y)>(p+1)^2
 \ee
in the region of outer communication $x>1$, $|y|\le1$.

\subsection{Ergosphere}

The solution has two Killing vectors $k=\partial_t$ and $m=\partial_\varphi$. As $x$
decreases, $k$ eventually becomes null on the ergosurface $f(x,y)=0$,
 \be
1-y^2=\sin^2\theta=\frac{p^2(x^2-1)^3}{4 q^2 x^4},
 \ee
marking the boundary of the ergosphere where $k$ is spacelike, $g_{tt}>0$.
On this ergosurface,
 \be\lb{gpp}
g_{\varphi\varphi} \equiv  F^{-1}\rho^2 - F\omega^ 2 =
\kappa^2\left[\frac{\Sigma}f(x^2-1)(1-y^2) - \frac{\Pi^2}{\Sigma
f}\right]
 \ee
is the difference between two terms which both diverge as $f\to 0$.
We show in Appendix A that these two poles cancel exactly so that
$g_{\varphi\varphi}$ is, as in the case of the Kerr metric, finite
and positive on the ergosurface.

As usual, inside the ergosphere the frame dragging effect
is manifest, forcing any neutral particle
to rotate with an angular velocity $\Omega$ in order that its world-line
$x^\mu(\tau)$ be time-like, i.e. $g_{\mu\nu}{\dot x^\mu}{\dot x^\nu}<0$.
This angular velocity must be within the bounds
 \be\lb{drag}
\Omega_-<\Omega<\Omega_+,\qquad \Omega_\pm=\frac{-g_{t\varphi} \pm
\sqrt{g_{t\varphi}^2-g_{tt} g_{\varphi\varphi}}}{g_{\varphi\varphi}}
= \frac{-F\omega \pm \rho}{g_{\varphi\varphi}}.
 \ee
In the case of the Kerr metric, the bounding velocities $\Omega_\pm$ approach each other
with decreasing $x$ till the horizon $\rho=0$, where $\Omega_-=\Omega_+$ is the angular velocity
of rotation of the horizon. In our case the situation appears to be different. When $x$
decreases towards $1$ with $y$ fixed ($|y|<1$), the various metric functions will behave as
$f = {\rm O}(\xi^{-2})$ ($f$ remaining negative), $\Sigma = {\rm O}(\xi^{-4})$,
$\Pi = {\rm O}(\xi^{-2})$, with $\xi^2 \equiv x^2-1$, so that from (\ref{gpp})
$g_{\varphi\varphi}$ will be dominated
by the constant negative first term. Therefore there is between the ergosphere and the
singularity $x=1$ a surface $g_{\varphi\varphi}(x,y)=0$ bounding the chronosphere
where $g_{\varphi\varphi}$ stays negative (see below). But, as in the case of the seed ZV2
one can also approach the singularity $x=1$ along a curve $(1-y^2)/(x^2-1)= X^2$ constant.
Then $f$, $\Sigma$ and $\Pi$ will go to constant values ($f$ remaining again negative),
so that now $g_{\varphi\varphi}$ will be dominated by the constant second term in (\ref{gpp})
and go to finite positive values, depending on the direction $X$, near the coordinate
singularities $x=1$, $y=\pm1$ which, as in the case of ZV2, are actually two disjoint
components of the degenerate horizon.

\subsection{Horizon}

Passing to the coordinates $X,\, Y$ via (\ref{XY}), (\ref{XY1}) one can see
that $Y=\pm1$ are degenerate (second order) horizons with $X$ being related to some angular
coordinates on them. Remarkably, while in the seed ZV2 metric the horizons were not
topological spheres, in the new solution they are, so the transformation ``improved''
the horizon geometry.  This indicates that the limits $q\to 0$ (ZV2 limit of the solution)
and $|Y|\to 1$ (near-horizon) do not commute. This can be seen by writing the function
$f$ in $(X,Y)$ coordinates as:
 \be
f = \frac1{(X^2+1)(X^2+Y^2)}\left[\frac{p^2}4(1-Y^2)^2 -q^2X^2(X^2+1)^2\right]\,.
 \ee
If we set first $q=0$ ($p=1$), $f$ develop a double zero $Y=\pm1$ corresponding to the double horizon
of the ZV2 degenerate static metric as in (\ref{ZVH}) while, for non-zero $q$, $f$ goes to
a non-zero limit $f_H$ for $|Y|\to1$.  We have
\ba\lb{fhor}
f_H(X) &=& -q^2X^2, \quad \Pi_H(X) = -q\lambda(p)X^2,\nn\\
\Sigma_H(X) &=& \frac{p\lambda(p)}2 + q^2(1+p)X^2 + \frac{q^4}4X^4,\quad e^{2\nu}= \frac4{p^2(1-Y^2)(X^2+1)^2}\,,
 \ea
with
 \be
\lambda(p) = \frac{(1+p)(8-4p+5p^2-p^3)}{2p} \ge 8
 \ee
(the lower bound being attained in the limit $q\to0$).

Inserting these behaviors in (\ref{gpp}), we find that $g_{\varphi\varphi}$
goes over to a positive function of $X$
 \be
g_{\varphi\varphi}\vert_H(X) = \frac{\kappa^2[\lambda(p)]^2X^2}{\Sigma_H(X)}
 \ee
including the case $q=0$ (the static ZV2 metric), in agreement with the result obtained in \cite{Kodama:2003ch}.
The Weyl coordinate $\rho = \kappa X(1-Y^ 2)/(X^2+Y^2)$ goes to zero for $Y\to\pm1$ so that, from
(\ref{drag}), the Killing vector $K =\partial_t-\Omega_{H } \partial_\varphi$ becomes null for $Y=\pm1$,
where the angular velocity of the two-component horizon is
 \be\lb{OmH}
\Omega_H = \lim_{|Y|\to 1} \Omega_\pm = \left. - \frac{\kappa\Pi}{\Sigma
g_{\varphi\varphi}}\right\vert_H = \left.
\frac{f}{\kappa\Pi}\right\vert_H = \frac{q}{\kappa\lambda(p)}.
\ee

In the horizon co-rotating frame $(\hat{t},X,Y,\p)$ defined by
$\hat{t} = t$, $\p = \varphi - \Omega_H t$, the two-dimensional sections
of the two horizon components $Y=\pm1$ have the same metric:
 \be\lb{methor}
ds_H^2  = \frac{4\kappa^2\Sigma(X)dX^2}{p^2(X^2+1)^4} +
\frac{\kappa^2\lambda^2(p)X^2d\p^2}{\Sigma(X)},
 \ee
Introducing a new angular coordinate $\eta$ by
 \be\lb{eta}
X = \tan(\eta/2) \quad (0 \le \eta \le \pi),
 \ee
(\ref{methor}) can be rewritten as
 \be\lb{methor1}
ds_H^2 = \frac{\kappa^2\lambda(p)}{2pl(\eta)}\left[d\eta^2 +
l^2(\eta)\sin^2\eta d\p^2\right],
 \ee
where
 \be
l(\eta) = \frac{p\lambda(p)}2\frac{(X^2+1)^2}{\Sigma(X)}
 \ee
is everywhere positive and finite. It follows that each horizon is
homeomorphic to $S^2$.

From (\ref{methor}) or (\ref{methor1}) we obtain the horizon area
 \be
{\cal A}_H = 4\pi\kappa^2\frac{\lambda(p)}{2p}.
 \ee
The corresponding areal radius is of the order of the total mass
$M$. More precisely, for small $q$, ${\cal A}_H \simeq
16\pi\kappa^2$, which leads to a {\em total} horizon area
$2{\cal A}_H(p=1) = 32\pi\kappa^2 = 8\pi M^2$, as in the static ZV2 case.
This is one-half of the horizon area for a Schwarzschild black hole
of the same asymptotic mass. The horizon area decreases monotonically with
increasing $|q|$ (decreasing $p$), until, for small $p$,
$\lambda (p) \sim 4/p$, leading to a total horizon area $2{\cal A}_H(p=0)
= 4\pi M^2$, which is again one-half of the horizon area for an
extreme Kerr black hole of the same asymptotic mass. It follows that
 \be
2{\cal A}_H(p) \le 8\pi M^2 \le
{\cal A}_{\rm Kerr} = 8\pi Mr_+,
  \ee
where $r_+ = M + \sqrt{M^2 - J^2/M^2}$ is the horizon radius for a Kerr black hole of
mass $M$ and angular momentum $J$, so that the total horizon area
of the present solution is always smaller than that for a Kerr black hole
of same mass and angular momentum. Let us also note that
it may seem surprising that the total horizon area in the limit
$p\to0$ is only one-half of that of the limiting solution, which is
the extreme Kerr black hole.

The metric (\ref{methor1}) has coordinate singularities at $\eta=0$
($X=0$), corresponding to the points where the two horizon components
intersect the regular semi-axes $\rho=0$, $|z|>\kappa$, and $\eta=\pi$
($X\to\infty$), corresponding to the ends of the interconnecting string.
From (\ref{fhor}) $l(0)=1$, so that the singularity at $\eta=0$ is spurious.
The curvature radius at $\eta=0$ is $\sqrt2$ times the areal horizon radius.
For $\eta=\pi$, we find
 \be\lb{alpha}
l(\pi) = \alpha(p) \equiv \frac{2p\lambda(p)}{q^4} > \frac8{q^4}> 8,
 \ee
meaning a conical singularity with negative deficit angle $2\pi(1-\alpha(p))$.

The evaluation of the electromagnetic functions $\v$
and $\Theta$ on the horizons leads to
 \be
\v_H(X) = \frac{q^2}4\left[p(2-p) - 2(1+p)X^2 - q^2X^4\right], \quad
\Theta_H(X) = \frac{q}4\left[\delta(p)X^2 + q^2\gamma(p)X^4\right],
 \ee
where
 \ba
\gamma(p) &=& \frac{(1 + p)(4 - p + p^2)}p = \lambda(p) +
\frac{q^2}2(2-p), \nn\\
\delta(p) &=& \frac{(1+p)^2(8-p^2+p^3)}p = 2(1+p)[\lambda(p) +
q^2(2-p)].
 \ea
Defining the electrostatic potential $\hat{v}$ in the static
near-horizon frame by $\hat{v}=v+\Omega_HA_\varphi$, we obtain
on the horizon
 \be
\hat{\v}_H(X) = \frac{q^2(2-p)}{2\lambda(p)}\Sigma_H(X).
 \ee
It follows that the horizon electromagnetic potential in the
co-rotating frame is, in the gauge $A(\infty)=0$,
 \be\lb{Ah}
\hat{A}_H= -\varepsilon\left(\frac{q^2(2-p)}{2\lambda(p)}\,dt +
\frac{\kappa q}4 \frac{(\delta(p)X^2+q^2\gamma(p)X^4)}{\Sigma(X)}\,
d\p\right).
 \ee

The vector potential (\ref{Ah}) generates a magnetic field
perpendicular to the horizon. Because the normals to the two
horizons $Y = 1$ and $Y = -1$ are oppositely oriented and the net
magnetic charge is zero, the magnetic lines of force must emerge
from one horizon and flow into the other horizon, so that the two
horizons can be considered as carrying exactly opposite magnetic
charges $P_+=-P_-=P_H$, where
 \be\lb{P}
P_H = \frac1{4\pi}\oint_{H_+}dA_{\varphi}\,d\varphi =
\varepsilon\frac{\kappa\gamma(p)}{2q}.
 \ee

\subsection{Chronosphere}

\begin{figure}[tb]
\begin{center}
\begin{minipage}[t]{0.42\linewidth}
\hbox to\linewidth{\hss%
  \includegraphics[width=1.2\linewidth,height=1.2\linewidth]{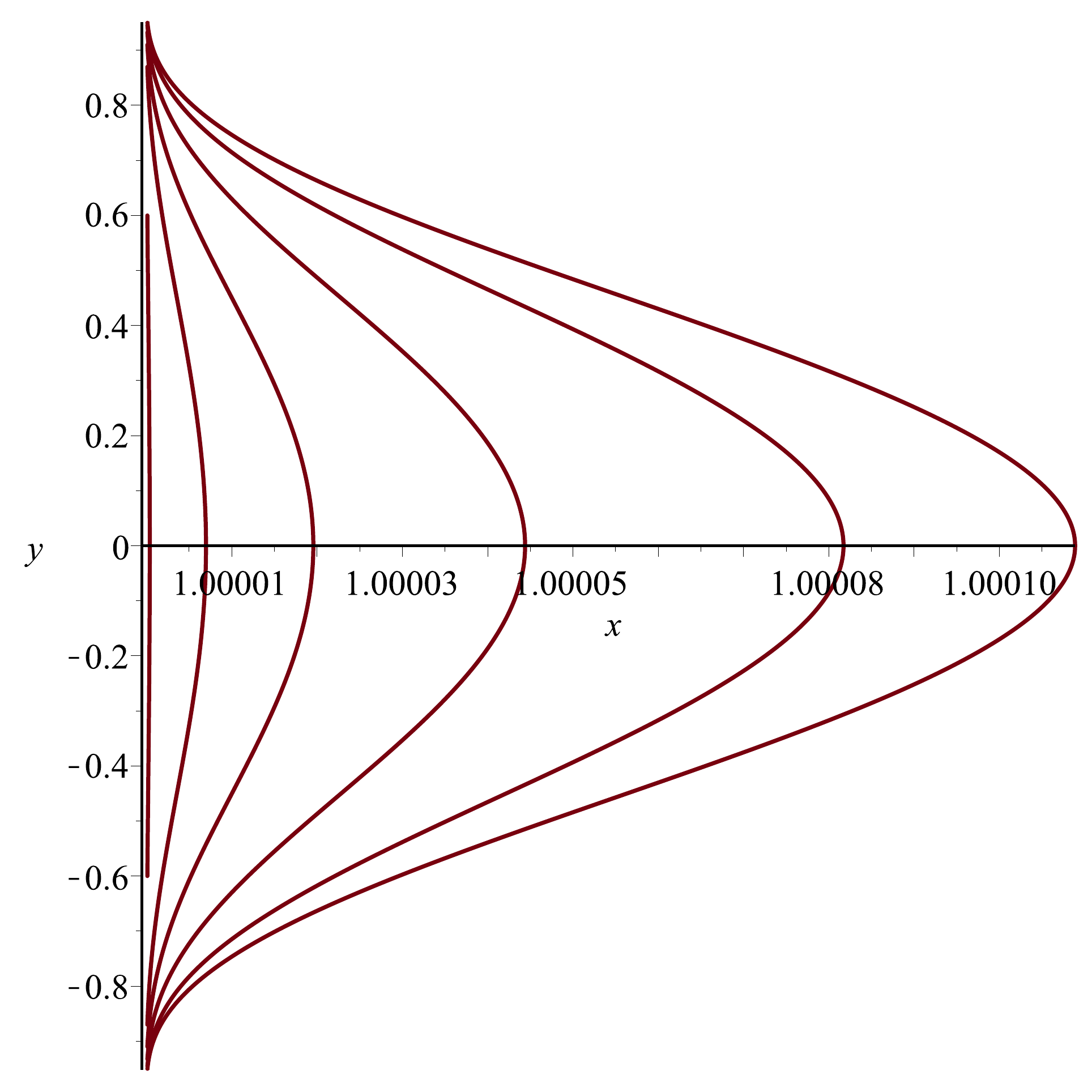}
\hss}
\end{minipage}
 \caption{\small Chronosphere boundary for
$q=.5,\,.6,\,.7,\,.8,\,.9,\,.9999$ (from left to right).} \label{F1}
\end{center}
\end{figure}

\begin{figure}[tb]
\begin{center}
\begin{minipage}[t]{0.42\linewidth}
\hbox to\linewidth{\hss%
  \includegraphics[width=1.2\linewidth,height=1.2\linewidth]{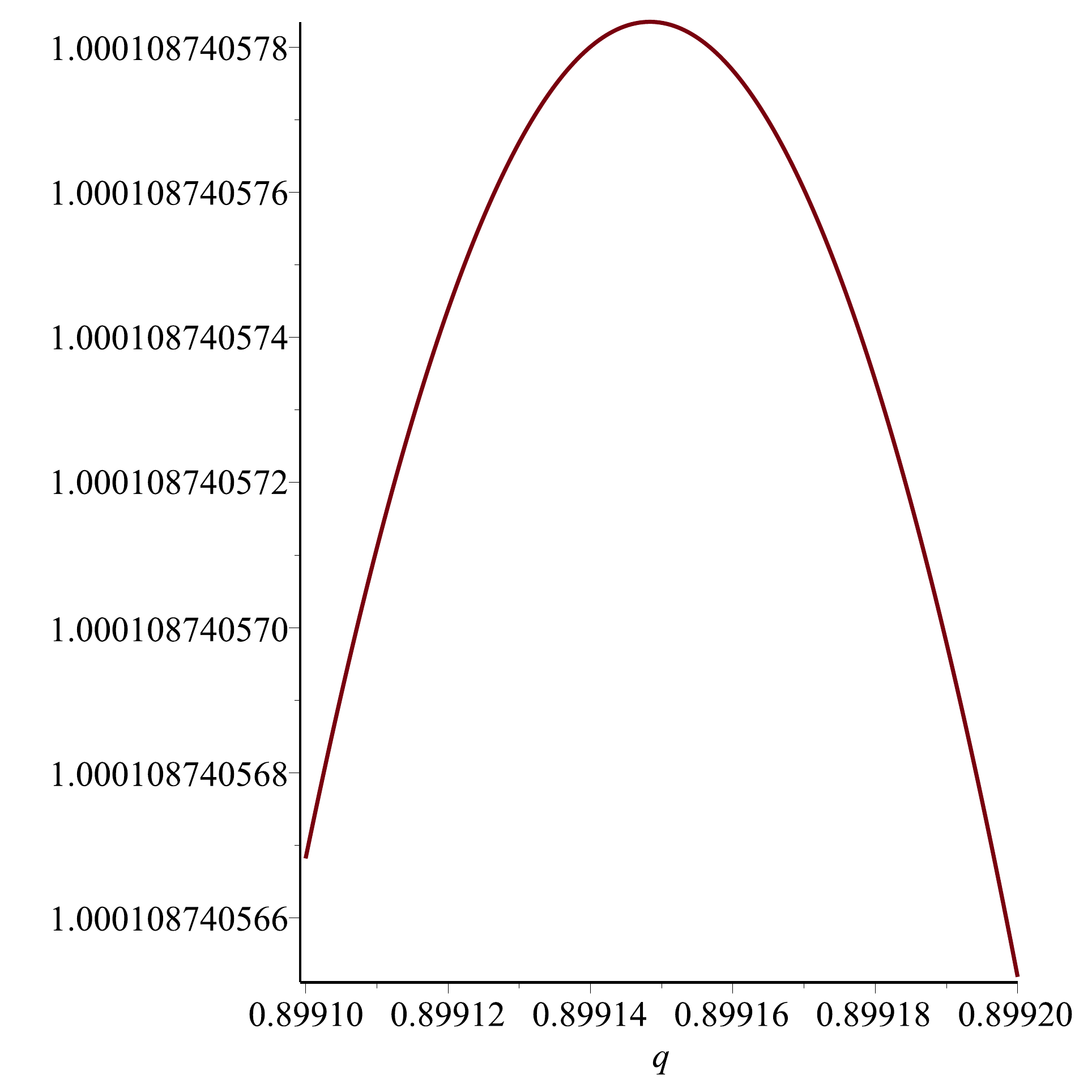}
\hss}
\end{minipage}
 \caption{\small Chronosphere boundary at the equatorial plane (maximal extension) in the vicinity of
the critical $q=q^*\approx .89915$ corresponding to the maximal size.} \label{F2}
\end{center}
\end{figure}

\begin{figure}[tb]
\begin{center}
\begin{minipage}[t]{0.35\linewidth}
\hbox to\linewidth{\hss%
  \includegraphics[width=1.2\linewidth,height=1.2\linewidth]{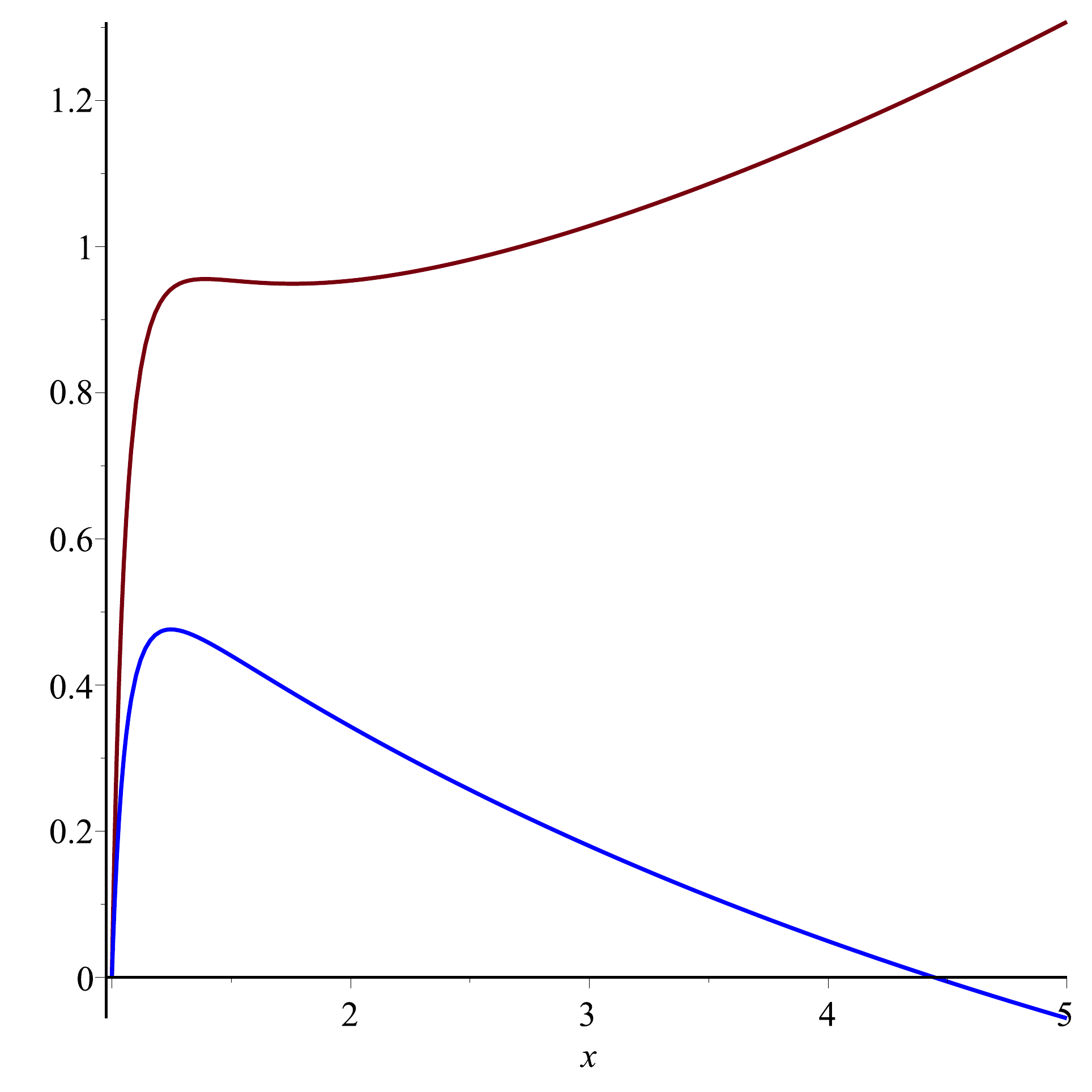}
\hss}
\end{minipage}
\caption{\small Plots of $10^{-2} g_{\varphi\varphi}$ (black) and $g_{tt}$ (blue) for $q=q^*$
at the equatorial plane. $g_{tt}$ changes sign to positive at the ergosphere boundary. Inside the
ergosphere   $g_{\varphi\varphi}$ seems positive too, but in fact it changes sign at the
chronosphere boundary, as can be seen with better resolution near $x=1$ (Fig. \ref{F4}).} \label{F3}
\end{center}
\end{figure}

\begin{figure}[tb]
\begin{center}
\begin{minipage}[t]{0.35\linewidth}
\hbox to\linewidth{\hss%
  \includegraphics[width=1.2\linewidth,height=1.2\linewidth]{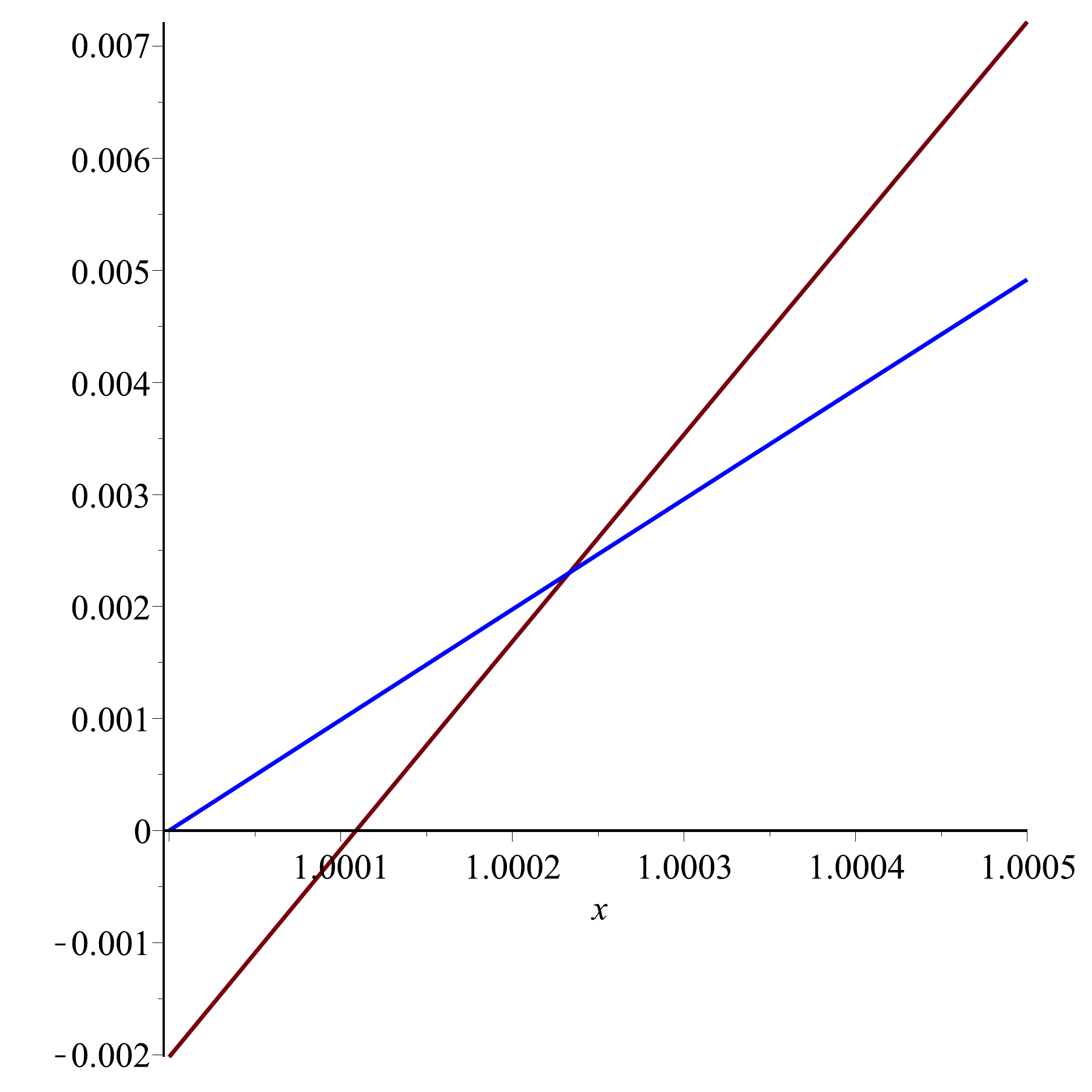}
\hss}
\end{minipage}
\caption{\small The blow up of the  plot Fig. \ref{F3} in the vicinity of $x=1$. Intersection
of zero by $g_{\varphi\varphi}$ (black)  marks the boundary of the chronosphere, inside which
$g_{tt}$ (blue) is still positive. Thus the chronosphere lies entirely inside the ergoregion. } \label{F4}
\end{center}
\end{figure}

We will call the region where the norm of the azimuthal Killing vector $m$ is negative,
$g_{\varphi\varphi}(x,y)<0$,
the chronosphere, since the time-like character of $\varphi\in [0, 2\pi]$ means that it contains
closed time-like curves. The boundary of the chronosphere (or causal boundary) $x=x_c(y)$ has its
maximal extension in the equatorial
plane $y=0$, and is a tiny region whose size in natural units $\kappa=1$ is less than $10^{-4}$.
In Fig. \ref{F1} we plot the family of curves $x=x_c(y)$ for different values of the
parameter $q$ (here assumed positive). The maximal size of the chronosphere is achieved
for $q=q^*\approx .89915$
as shown in  Fig. \ref{F2} where $x_c(y=0)$ is plotted as function of $q$ in the vicinity of $q^*$.
The plot of $g_{\varphi\varphi}$ (factored by $10^{-2}$ for compatibility) in the equatorial
plane for $q=q^*$ is shown on Figs. \ref{F3}, \ref{F4} together with $g_{tt}$.
The Fig. \ref{F4} shows a simple zero of $g_{\varphi\varphi}$
at $x=x_c$, while $g_{tt}$ remains positive, as it has to be inside the ergosphere.

\begin{figure}[ht]
%\centering
\includegraphics[scale=0.4]{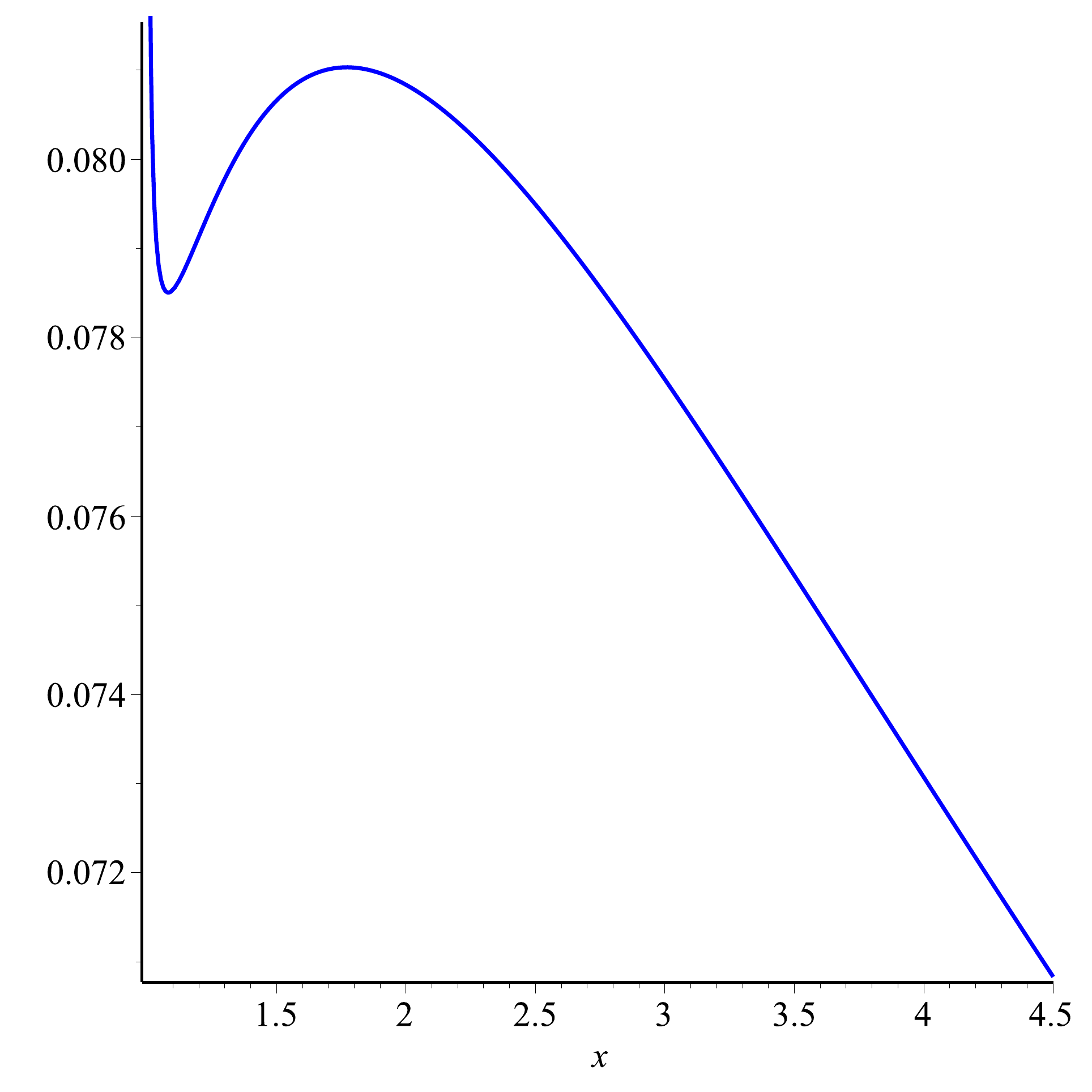}
\includegraphics[scale=0.4]{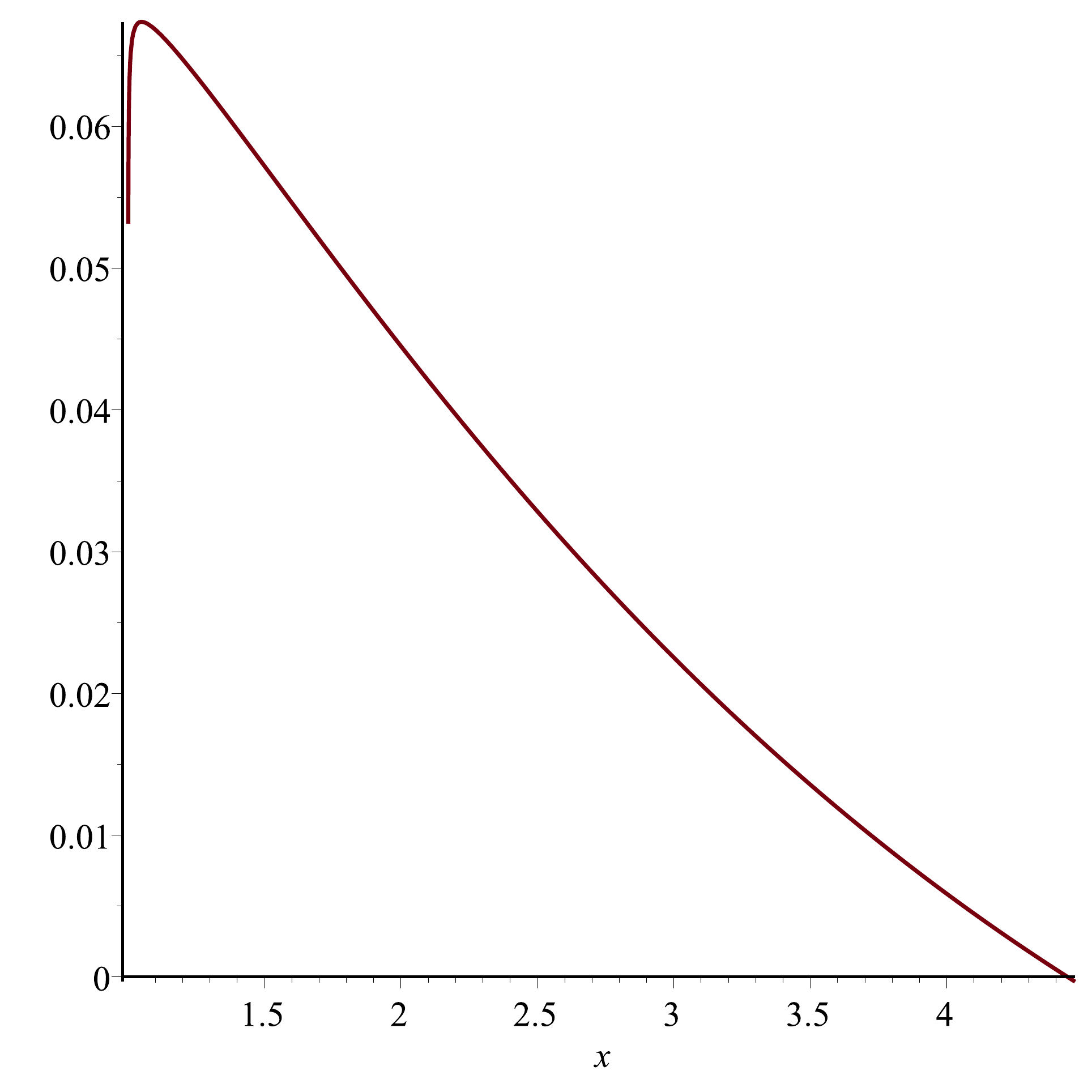}
\caption{Bounds of dragging angular velocities  in  the ergosphere outside the chronosphere at the equatorial plane: $\Omega_+$ (left panel) and  $\Omega_-$ (right panel).}
\label{F5}
\end{figure}

The  frame dragging velocities $\Omega_\pm$ are plotted for $y=0$ and $q=q^*$ in Fig. \ref{F5} 
as  functions of $x$ in the ergosphere  outside the chronosphere. Both are positive there. $\Omega_+$
diverges at the chronosphere boundary as $x\to x_c+0$ while $\Omega_-$ remains bounded,
\be
\lim_{x\to x_c-0}\Omega_-(x)=\frac{g_{tt}}{2\left|g_{t\varphi}\right|}
\Big|_{x=x_c}=\frac1{2\omega}\Big|_{x=x_c}.
\ee
Inside the chronosphere, $\Omega_+\to -\infty$ as $x\to x_c-0$, so that now $\Omega_+<\Omega_-$.

Thus the conditions for an observer's world-line to remain time-like inside the chronosphere are
somewhat similar to (\ref{drag}), but with an ``and'' replaced by an ``or'',
\be\lb{drag1}
\Omega<\Omega_+\,,\quad{\rm or} \quad \Omega>\Omega_-\,.
\ee

We expect that, similarly to the cases studied in \cite{RTN,NW}, all possible closed time-like curves
inside the chronosphere can be shown to be non-geodesic. One can also argue that typical
quantum effects would be expected to be of the order of $1/\kappa$, i.e. one in the above units.
The classical chronosphere having a size four orders of magnitude smaller would then be far
outside the validity of the classical theory. This can be contrasted with the case of the Taub-NUT metric,
where the chronosphere around the Misner string in non-compact, and whose characteristic
size is of the order of the horizon radius.

\begin{figure}[tb]
\begin{center}
\begin{minipage}[t]{0.58\linewidth}
\hbox to\linewidth{\hss%
  \includegraphics[width=1.5\linewidth,height=1.2\linewidth]{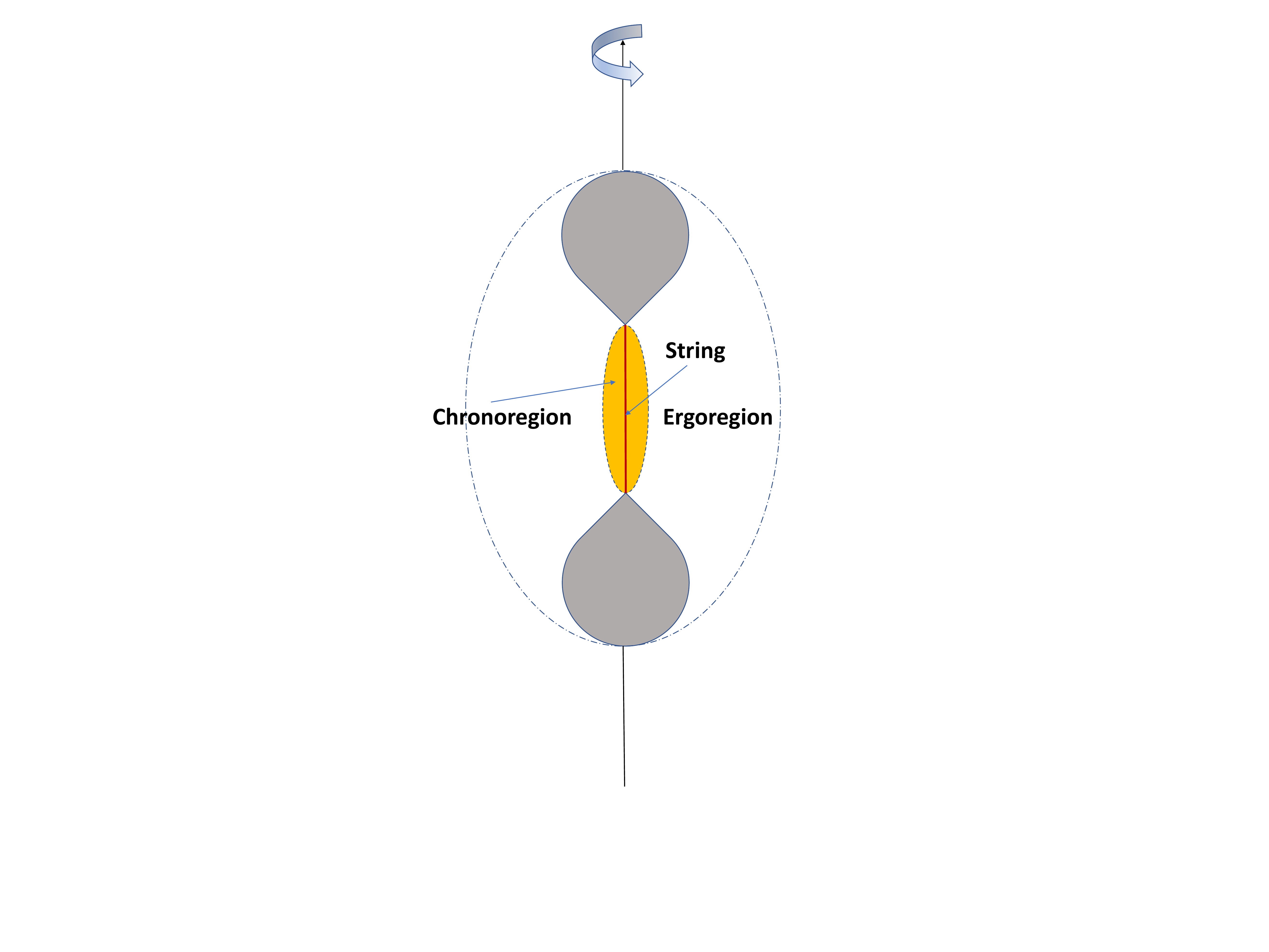}
\hss}
\end{minipage}
 \caption{\small Relative positions of the ergosphere, chronosphere, constituent black holes  and string singularity.} \label{F7}
\end{center}
\end{figure}

The relative positions of the ergosphere, chronosphere, horizons and string singularity are shown in Fig. \ref{F7}.
%%%%%%%%%%%%%%%%%%%%%%%%%%%%%%%
In the Kerr metric the chronosphere exists too but it lies  inside the event horizon, and thus
is ignored. In the case of the TS2 solution, closer to ours, the chronosphere is not inside
the ergosphere. The ergoregion there has an inner boundary within which sits the chronosphere.
The two surfaces intersect on a singular ring, which is a strong curvature singularity \cite{Kodama:2003ch}.

\subsection{Singular string}

This is the segment $x=1$, $-1<y<1$ ($\rho=0$, $-\kappa<z<\kappa$) between the two horizons.
For $\xi^2 \equiv x^2-1\to0$ (with $y^2<1$), the solution
(\ref{anmet})-(\ref{anem}) reduces to:
 \ba
ds^2 &\sim& - \frac{\kappa^2q^2}4(1-y^2)^2\,d\varphi^2 +
\frac{\kappa^2q^4}{p^2(1-y^2)}\left[\frac{dy^2}{1-y^2}\right. \nn\\
&& \left. + d\xi^2 + \frac{4p^2}{\kappa^2q^6}\xi^2\left(dt -
\kappa\left(\frac{\lambda(p)}q+q(1-p/2)(1-y^2)\right)
d\varphi\right)^2\right] \lb{rodmet}\\
A &\sim&
\varepsilon\left[\left(1-\frac{2(1+p)\xi^2}{q^2(1-y^2)}\right)dt -
\kappa\left(\frac{\gamma(p)}{q}
+\frac{q(1-y^2)}2\right)d\varphi\right], \lb{rodA}
 \ea
where we have neglected irrelevant terms of order $\xi^2$ and
higher.

The singularity at $\xi=0$ looks like a conical singularity. It would be one
if the time coordinate $t$ was periodic with period $T$, and it would disappear altogether
for the value of the period $T = \pi\kappa q^3/p$. The electromagnetic invariants
 \be
\frac12 F^{\mu\nu}F_{\mu\nu} \sim \frac{4p^2}{\kappa^2q^6}[(1+p)^2 -
q^2y^2], \quad \frac1{4\sqrt{|g|}}\varepsilon^{\mu\nu\rho\sigma}
F_{\mu\nu}F_{\rho\sigma }\sim - \frac{8p^2(1+p)}{\kappa^2q^5}\,y
 \ee
are finite, as well as the mixed Ricci tensor components, the
diagonal components behaving as
 \be
R^t_t \sim R^x_x \sim -R^\varphi_\varphi \sim -R^y_y \sim
\frac{4p^2}{\kappa^2q^6}[(1+p)^2 + q^2y^2],
 \ee
with $R^\varphi_t \sim R^x_y \sim 0$, so that the Ricci square scalar
 \be
R^{\mu\nu}R_{\mu\nu} \sim \frac{64p^4}{\kappa^4q^{12}}[(1+p)^2 +
q^2y^2]^2
 \ee
stays finite near the singularity.
On the string $x=1$ both dragging velocities  $\Omega_\pm$  tend to zero in accordance with the law
\be
\lim_{x\to 1}\Omega_\mp\sim \pm\frac {4}{q^2}\frac{(x^2-1)^{1/2}}{\kappa (1-y^2)^{3/2}}
\ee
It follows from (\ref{drag1}) that in the near-string limit $x\to1$ all observer angular velocities are allowed,
which is the exact opposite of the near-horizon limit, so in this sense the interconnecting string
can be viewed as an anti-horizon.

The string geometry becomes more transparent in the horizon co-rotating frame $(t,x,y,\p)$, with
$d\p = d\varphi - \Omega_Hdt$. The near-string metric (\ref{rodmet}) transforms to
 \ba\lb{rodmet1}
ds^2 &\sim&
q^4\left[-\frac{(1-y^2)^2}{4\lambda^2(p)}\left(dt+\Omega_H^{-1}
\,d\p\right)^2\right. \nn\\
&+& \left.\frac{\kappa^2}{p^2(1-y^2)}\left(\frac{dy^2}{1-y^2} +
{d\xi^2} + \alpha^2\xi^2\,d\p^2\right)\right].
 \ea
We recognize in (\ref{rodmet1}) the metric of a spinning cosmic string \cite{DJH,GC85} in a
curved spacetime, with (negative) tension per unit length $(1-\alpha)/4$, where $\alpha(p)$
is given in (\ref{alpha}),
and ``spin'' $\Omega_H^{-1}/4$, where $\Omega_H(p)$ is given in (\ref{OmH}).
In view of the fact that the finite-length string connects two black
holes, this spin should actually be interpreted as a gravimagnetic
flow along the Misner string connecting two opposite NUT sources at
$\rho=0$, $z = \pm\kappa$, with the gravimagnetic potential $\omega/2 =
N_+\cos\theta_+ + N_-\cos\theta_-$ where $\cos\theta_\pm=\mp1$   along
the string, and $N_+=-N_-=N_H$, where
 \be\lb{N}
N_H = \dfrac{\kappa\lambda(p)}{4q}.
 \ee

Similarly, the constant
contribution $-\varepsilon\kappa\gamma(p)/q$ to $A_\varphi$ in
(\ref{rodA}) should be interpreted as the magnetic flow along a
Dirac string connecting two opposite monopoles at $z=\pm\kappa$ with
magnetic charges $P_+=-P_-=P_H$, where $P_H$ is the horizon magnetic
charge already given in (\ref{P}). The non-constant contribution
gives rise to a magnetic field density $\sqrt{|g|}B^\xi =
F_{y\varphi} = \varepsilon\kappa qy$, which leads to an intrinsic
string magnetic moment
 \be
\mu_S = \frac1{4\pi}\int_{-1}^{+1}\sqrt{|g|}B^\xi\,z\,2\pi\,dy =
\frac{\varepsilon\kappa^2q}3 = \frac\mu3
 \ee
(to obtain the total magnetic moment $\mu$, the magnetic dipole
contribution $2\kappa P_H$ and the sum of the horizon magnetic
moments should be added to this). Other string observables (mass,
angular momentum, electric charge) shall be evaluated in the next section.

\setcounter{equation}{0}
\section{Komar-Tomimatsu observables}\lb{observables}

Tomimatsu has shown \cite{tom84} that, by using the Ostrogradsky theorem and thef
Einstein-Maxwell equations, the Komar mass and angular momentum at infinity
 \be
M = \frac1{4\pi}\oint_\infty D^\nu k^{\mu}d\Sigma_{\mu\nu}, \quad J
= -(1/8\pi)\oint_\infty D^\nu m^{\mu}d\Sigma_{\mu\nu}
 \ee
($k^\mu = \delta^\mu_t$, $m^\mu = \delta^\mu_\varphi$) can be
transformed into the sums over the boundary surfaces $S_n$
(here, the two horizons and the string) $M = \sum_n M_n$, $J =
\sum_n J_n$, with
 \ba\lb{MJn}
M_n &=&
\frac1{8\pi}\oint_{S_n}\left[g^{ij}g^{ta}\partial_jg_{ta}
+2(A_t F^{it}-A_\varphi F^{i\varphi})\right]d\Sigma_i. \nn\\
J_n &=& -\frac1{16\pi}\oint_{S_n}\left[g^{ij}g^{ta}
\partial_jg_{\varphi a} +4A_\varphi F^{it}\right]d\Sigma_i.
 \ea

In the case of {\em rotating} black holes, the horizon Komar
mass and angular momentum ((\ref{MJn}) with $S_n=H$)
reduce on the horizons to \cite{tom84,smarr}
 \ba
M_H &=& \frac1{8\pi}\oint_H\left[\omega\partial_z{\rm Im}\E +
2\partial_z(A_\varphi\,{\rm Im}\psi)\right]dz\,d\varphi,\lb{MH}\\
J_H &=& \frac1{8\pi}\oint_H\omega\left[-1 + \dfrac12\omega\,\partial_z{\rm Im}\E
+ \partial_z(A_\varphi\,{\rm Im}\psi) + \omega\hat{A}_t\partial_z{\rm
Im}\psi\right]dz\,d\varphi \lb{JH}
 \ea
(the second term in (\ref{MH}) was omitted in \cite{tom84}).

Transforming from the Weyl coordinates $\rho,z$ to the coordinates $X,Y$, and taking
into account the constancy of $\omega$ and $\hat{A}_t$ over the horizon,
the expressions (\ref{MH}) and (\ref{JH}) can be integrated to\footnote{In the
present case the (degenerate) horizons are pointlike in coordinates $(\rho,z)$,
so that the first term in the integrand of (\ref{JH}) does not contribute to $J_H$.}
 \ba
M_H &=& \frac14\bigg[\omega{\rm Im}\E + 2A_\varphi\,{\rm Im}\psi\bigg]_{X=\infty}^{X=0},\\
J_H &=& \frac14\omega\left[\dfrac12\omega{\rm Im}\E + A_\varphi\,{\rm Im}\psi +
\omega\hat{A}_t{\rm Im}\psi\right]_{X=\infty}^{X=0},
 \ea
evaluated over the upper horizon $Y=1$ (the contributions of the two horizons are equal).
Using $\omega_H=1/\Omega_H$, we find
 \ba
& M_{H_\pm} =& \frac\kappa{p} +\frac{\kappa p}2,\lb{MH1}\\
&J_{H_\pm} =&
\frac{\kappa^2}{8qp}\left[2\lambda(p)(2+p^2)+q^2p(1+p)(2-p)\right].\lb{JH1}
 \ea
The horizon mass (\ref{MH1}) is larger than half of the global mass
$2\kappa/p$, so the string must have negative mass.

Similarly, the horizon electric charge
 \be
Q_H = \frac1{4\pi}\int_H\sqrt{|g|}F^{t\rho}dzd\varphi
 \ee
may be transformed into the Tomimatsu integral \cite{tom84}
 \be
Q_H = -\dfrac1{4\pi}\oint_H\omega\,d\,{\rm Im}\psi\,d\varphi,
 \ee
leading to
 \be\lb{QH}
Q_{H_\pm} = -\dfrac{\varepsilon\kappa(1+p)}2.
 \ee
To ensure global electric neutrality, the
string must be also charged, which we shall now check.

The near-string covariant component
$F_{t\xi}$ of the radial electric field vanishes to order ${\rm
O}(\xi)$, but on account of $g_{tt} = {\rm O}(\xi^2)$ and
$\sqrt{|g|} = {\rm O}(\xi)$, the radial electric field density
$\sqrt{|g|}F^{t\xi}$ is finite and constant along (a small 
cylinder centered on) the string, leading to the electric charge
 \be
Q_S = \frac1{4\pi}\int_{-1}^{+1}\sqrt{|g|}F^{t\xi}\,2\pi\,dy =
\varepsilon\kappa(1+p).
 \ee
This string electric charge together with the horizon electric
charges lead to a vanishing total electric charge
 \be
Q_{H_+} + Q_{H_-} + Q_S = 0,
 \ee
a vanishing electric dipole moment, and a contribution to the total
electric quadrupole moment, to which must be added that of the two
opposite horizon electric dipole moments generated by the rotation
of the horizon magnetic charges, and the sum of the horizon electric
quadrupole moments.

The string mass and angular momentum can be evaluated from
(\ref{MJn}) integrated over a small cylinder centered on 
the string $x=1$, $-1<y<1$, and are the
sum of gravitational and electromagnetic contributions. Although, in
the co-rotating frame, the string is a spinning cosmic string with
negative tension, and thus presumably negative gravitational mass,
in the global frame the gravitational contribution to the string
mass is -- surprisingly -- positive. However it is overwhelmed by
the negative electromagnetic contribution $-Q_SA_t(\xi=0)$,
resulting in a net negative string mass
 \be
M_S = \kappa - \kappa(1+p) = - \kappa p,
 \ee
which represents the binding energy between the two black holes of
mass $(\kappa/p + \kappa p/2)$, leading to the total mass
 \be
M = M_{H_+} + M_{H_-} + M_S = \frac{2\kappa}p.
 \ee
The fact that the string mass is negative explains the repulsion
experienced by test particles in geodesic motion near the string
(antigravity).

Similarly, the string angular momentum is the sum of gravitational
and electromagnetic contributions
 \ba\lb{JS}
J_S &=&
\kappa^2\left[\frac{\lambda(p)}{2q}+\frac{q}3\left(1-\frac{p}2
\right)\right] -
\kappa^2(1+p)\left[\frac{\gamma(p)}{q}+\frac{q}3\right]\nn\\
&=& \frac{\kappa^2}{2q}\left[\lambda(p) - 2(1+p)\gamma(p) -
pq^2\right].
 \ea
The first term $\kappa^2\lambda(p)/2q$ is the NUT dipole $2\kappa
N_H$. The second term can be understood as the charge-monopole
angular momentum contribution $Q_S(P_{H_+}-P_{H_-})$. If this interpretation
is correct, the remainder $-\kappa pq/2$ corresponds to the
intrinsic string angular momentum. It can be checked that the
horizon angular momenta (\ref{JH1}) and the total string angular
momentum (\ref{JS}) add up to the net angular momentum (\ref{asMJ}):
 \be
J = J_{H+} + J_{H_-} + J_S = \frac{\kappa^2 q(4+p^2)}{p^2}.
 \ee

\setcounter{equation}{0}
\section{Geodesics}
Two obvious first integrals of the geodesic equations of motion are
 \ba
&& F\left(\dot{t}-\omega\dot{\varphi}\right) = E,\\
&& F^{-1}\rho^2\dot{\varphi} + F\omega
\left(\dot{t}-\omega\dot{\varphi}\right) = L,
 \ea
where $\dot= d/d\tau$, and $E$ (energy) and $L$ (orbital angular
momentum) are two constants of the motion. A third first integral is
 \be\lb{geo1}
\dfrac{ds^2}{d\tau^2} \equiv g_{\mu\nu}\dot{x}^\mu\dot{x}^\nu \equiv
T+U=\epsilon,
 \ee
where $\epsilon=-1$, $0$ or $+1$ for timelike, null or spacelike
geodesics, and
 \ba
T &=& \kappa^2\Sigma e^{2\nu}\left(\dfrac{\dot{x}^2}{x^2-1} +
\dfrac{\dot{y}^2}{1-y^2}\right) > 0,\lb{T}\\
U &=& \dfrac{(L-E\omega)^2F}{\rho^2} - \dfrac{E^2}F. \lb{U}
 \ea
The fourth equation is the geodesic equation for the coordinate $y$,
which reads:
 \ba\lb{geoy}
&& \frac{2\kappa^2}{\sqrt{1-y^2}} \frac{d}{d\tau}\left(\frac{\Sigma
e^{2\nu}\dot{y}}{\sqrt{1-y^2}}\right) =
\frac{E^2\Sigma}f\left[2\frac{\Sigma_y}\Sigma+2\nu_y -
\frac{f_y}f\right] + \frac{2\kappa E\Pi}f\left[\frac{\Pi_y}\Pi -
\frac{f_y}f\right]  \nn\\ && + \epsilon\left[\frac{\Sigma_y}\Sigma +
2\nu_y\right] - \frac{(Lf-\kappa E\Pi)^2}{\kappa^2 f\Sigma
(x^2-1)(1-y^2)}\left[\frac{f_y}f + 2\frac{y}{1-y^2} + 2\nu_y\right],
 \ea
where $\rho^2$ is given by Eq. (\ref{sphero}),
$F_y=\partial F/\partial y$ and so on.

Contrary to the Kerr case,
there is no Carter constant corresponding to the second order Killing tensor,
so the system of $(x(\tau),\,y(\tau))$ equations cannot be decoupled.
However one can derive a separate non-linear differential equation for
the function $y(x)$ describing the geodesic trajectories in the $(x,\,y)$ plane.
In the ZV2 case such an equation was given in \cite{Kodama:2003ch}.  To this aim one writes
$\dot{y}=y'\dot{x}$, where $y'=dy/dx$, and substitute this in the Eqs. (\ref{geo1}, \ref{T}):
\be
 \kappa^2\Sigma e^{2\nu}{\dot{x}^2}\left(\dfrac1{x^2-1} +
\dfrac{ {y'}^2}{1-y^2}\right)+U=\epsilon\,,
\ee
to express $\dot{x}^2$  as a function of three variables $\chi(x,y,y')$:
\be
\dot{x}^2 = \chi \equiv \frac{(\epsilon-U)(x^2-1)(1-y^2)}{\kappa^2 \Sigma e^{2\nu}\left[1-y^2+(x^2-1)y'^2\right] }\,.
\ee
Making the same substitution in Eq. (\ref{geoy}) we obtain the desired equation
 \be
y''\left(\chi+y'\chi_{y'}/2 \right)+y'(\partial_x+y'\partial_y)\ln\left(\Sigma e^{2\nu}\chi^{1/2}\right)+2 y y'^2\chi(1-y^2)^{-1}=\Phi(1-y^2)/2\kappa^2\,.
 \ee

Let us first discuss the behavior of geodesics near the string
$S$ ($x=1$, $y^2<1$). We have seen that near $S$ both $F^{-1}\rho^2$
and $\omega$ go to finite limits depending on $y$, so that the first
centrifugal term in $U$ is negative and bounded, while from
(\ref{rodmet}) the second term $-E^2/F$ is positive and increases
without bound, so that the geodesics are reflected by a potential
barrier. This argument breaks down in the exceptional case $E=0$,
where only the (attractive) centrifugal potential remains, so that
these geodesics terminate (or originate) on the singularity $S$.
However, the timelike or null geodesics ($\epsilon\le0$) are
confined to the region where the centrifugal potential is
attractive, i.e. inside the ergosphere, and the orbits must have a
turning point somewhere, and by reason of symmetry end again on the
singularity.

Consider now geodesics approaching either of the two points $H_\pm$
($x=1$, $y=\pm1$). The simplest case is that of axial geodesics
$\rho²=0$. A first possibility is $y=\pm1$ (axial geodesics originating
from infinity), which necessitates $L=0$, and in which case
$\omega=0$. Then $fe^{2\nu}=x^2-1$, so that Eq. (\ref{geo1}) reduces
to the exact equation
 \be\lb{geo2}
\dot{x}^2 - \frac{\epsilon p^2(x^2-1)^2}{4\kappa^2x^2\Sigma(x)} =
\frac{E^2}{\kappa^2},
 \ee
where $\Sigma(x)$ goes to a finite limit for $x\to 1$ . Clearly
these special geodesics attain $x=1$ in a finite affine time, with
an affine velocity equal to that of light, and can be analytically
continued to $x<1$, all the way to $x \to -\infty$ (from (\ref{Sig})
$x^2\Sigma(x)$ admits the lower bound $p^2/2$). The other possibility
is $x=1$ (geodesics along the string from one horizon to the other).
Then $F=0$, so finiteness of $U$ requires $E=0$, the first integrated
geodesic equation
 \be\lb{geostring}
\dot{y}^2 - \frac{\epsilon p^2}{\kappa^2q^4}(1-y^2)^2 = \frac{4L^2
p^2}{\kappa^4q^6}
 \ee
showing that the two horizon components are connected by axial
spacelike geodesics, null geodesics with $L\neq0$, as well as
timelike geodesics with $|L|>\kappa q/2$.

In the generic case, let us assume that the geodesic
hits the point $H_\pm$ tangentially to the curve
 \be\lb{X}
1-y^2 = X^2(x^2-1),
 \ee
i.e. that the initial conditions for the orbit $y(x)$ at $x=1$ are
$y(1)=\pm1$, $y'(1)=\mp X^2$, and show that these conditions are
consistent with the geodesic equation for $y$ (\ref{geoy}). First we
observe that, as $x^2\to1$, the functions $f$, $\Sigma$ and $\Pi$
all go to finite limits (given below in (\ref{fhor})), while the
logarithmic derivatives $\Sigma_y/\Sigma$, $\nu_y$ and $f_y/f$ are
all of order $1/(x^2-1)$. It follows that the right-hand side of
(\ref{geoy}) is dominated by the last term. Furthermore, $f \sim
-q^2x^2(1-y^2)/(x^2-1)$, so that to leading order $f/(1-y^2)$ does
not depend on $y$, and the square bracket in this last term is
dominated by the term $2\nu_y$. Accordingly, near $x=1$ Eq.
(\ref{geoy}) may be replaced to leading order by
 \be\lb{geoy1}
\frac{2\kappa^2}{\sqrt{1-y^2}} \left(\frac{\Sigma
e^{2\nu}\dot{y}}{\sqrt{1-y^2}}\right)^{\dot{}} \simeq
-2\dfrac{(L-E\omega)^2F}{\rho^2}\nu_y \simeq -2U\nu_y \simeq
2T\nu_y,
 \ee
where we have used (\ref{U}) and (\ref{geo1}) to leading order.
Because $\Sigma$ and $\dot{y}$ go to finite limits for $x\to1$,
their derivatives may be neglected in (\ref{geoy1}), which may be
again replaced by
 \be
2\kappa^2\Sigma e^{2\nu}\,\frac{\dot{y}}{1-y^2}\left[2\dot\nu +
\frac{y\dot{y}}{1-y^2}\right] \simeq 2T\nu_y.
 \ee
Comparing with the definition (\ref{T}) of $T$, we arrive at the
equation
 \be
- 2(x^2-1)(\nu_x-X^2y\nu_y) + 1 \simeq (x^2-1)(1+X^2)y\nu_y,
 \ee
which, using the partial derivatives of (\ref{e2nu})
 \be
\nu_x \simeq \left(2-\frac3{1+X^2}\right)\frac1{x^2-1}, \qquad \nu_y
\simeq \frac{3y}{1+X^2}\frac1{x^2-1},
 \ee
is seen to be satisfied. This shows that the behavior (\ref{e2nu})
of the metric function $e^{2\nu}$, inherited from the ZV2 solution,
is essential for the consistency of the assumption (\ref{X}).

Considering now the effective radial equation $T \sim -U$ where
$y(x)$ is given by (\ref{X}) and $U$ is dominated by the negative
centrifugal contribution, and using the limit
 \be\lb{lime2nu}
\lim_{x\to1}{\dfrac{e^{2\nu}\rho^2}{x^2-1}} =
\dfrac{4\kappa^2}{p^2}\,\dfrac{X^2}{(1+X^2)^3},
 \ee
we see that the `radial' velocity $\dot{x}$ goes for  $x\to1$
to a finite limit
 \be\lb{radvel}
\dot{x}_H = -
\frac{p(qL+\kappa\lambda(p)E)}{2\kappa^2}\,\frac{1+X^2}{\Sigma_H(X)}.
 \ee
The axial radial velocity $-E/\kappa$ is recovered in the limit ($X\to0$, $L\to0$).

The geodesic equations being analytical in $x$ and $y$, these geodesics
can be smoothly continued through the horizon $x=\pm y=1$ to an
interior region with $x<1$ and $y^2>1$, without changing the
signature of the metric because the simultaneous sign change of
$x^2-1$ and $1-y^2$ leads to a sign change of $e^{2\nu}$,
proportional to $x^2-y^2$.

\setcounter{equation}{0}
\section{Beyond the horizons}
The metric inside the black hole (region $II$) is again given by
(\ref{anmet}) where now $-1<x<1$ and $y\ge1$ (North interior region
$II_+$) or $y\le-1$ (South interior region $II_-$). These two
isometrical interior regions are actually disconnected, each being
bounded by two horizons $H_\pm$ ($x=1,\,y=\pm1$) and $H'_\pm$
($x=-1,\,y=\pm1$). Behind the second horizons $H'_\pm$ lie two new
exterior regions with $x<-1$ and $-1\le y\le1$. In the most
economical maximal analytic extension, these two isometrical regions
can be identified (exterior region $III$).

In region $II$ the role of the radial coordinate is now played by
$y$, $x$ being related to the natural angular coordinate by
$x=\cos\theta$. It follows that the coordinate singularity at
($x=1$, $|y|>1$) as well as that at ($x=-1$, $|y|>1$) are axial
singularities of the cosmic string type, the near-singularity metric
and electromagnetic potential, obtained by taking $1-x^2 = \xi^2\to0$
being now
 \ba
ds^2 &\sim& - \frac{\kappa^2q^2}4(y^2-1)^2\,d\varphi^2 +
\frac{\kappa^2q^4}{p^2(y^2-1)}\left[\frac{dy^2}{y^2-1}\right. \nn\\
&& \left. + d\xi^2 + \alpha^2\Omega_H^2\xi^2\left(dt -
\kappa\left(\frac{\lambda(\pm p)}q-q(1\mp p/2)(y^2-1)\right)
d\varphi\right)^2\right] \lb{rodmet2}\\
A &\sim&
\varepsilon\left[\left(1-\frac{2(1\pm p)\xi^2}{q^2(y^2-1)}\right)dt -
\kappa\left(\frac{\gamma(\pm p)}{q}
-\frac{q(y^2-1)}2\right)d\varphi\right], \lb{rodA2}
 \ea
These two strings have different tensions $(1-\alpha(\pm p))/4$ and
different spins $\Omega_H(\pm p)^{-1}/4$ (it is clear from
(\ref{kin}) that the exchange $x\to-x$ is equivalent to the exchange
$p\to-p$), $\alpha(\pm p)$ being given in (\ref{alpha}) and
$\Omega_H(\pm p)$ in (\ref{OmH}) (with the product $\alpha(\pm
p)\Omega_H(\pm p) = \mp2p/\kappa q^3$). As in the case of Sect. 4,
the effective potential of (\ref{geo1}) increases for $E\neq0$ as
$1/\xi^2$, so that no geodesics can reach these singular strings,
except for exceptional geodesics with $E=0$. Again, these cosmic
strings are themselves geodesic with $E=0$ and the first integrated
geodesic equation (\ref{geostring}).

There is also an apparent singularity at $x=0$. However evaluation
of the various metric elements near $x=0$ leads to the regular
behavior
 \ba
ds^2 &\sim& - \frac1{1+q^2y^2}\left[dt+\frac{4\kappa
q}p\left(1+\frac{q^2}8(y^2-1)\right)(y^2-1)d\varphi\right]^2 \nn\\
&+& \kappa^2(1+q^2y^2)\left[y^{-6}\left(dx^2 +
\frac{dy^2}{y^2-1}\right) + (y^2-1)d\varphi^2\right].
 \ea
Again, there is no ring singularity in region $II$, because $\Sigma$
is the sum of two squares, the second of which can vanish only for
$x = p/2$, and it is then easy to show that $\Sigma(p/2,y)
> 4$.

The metric is by construction regular for $y=\pm1$, being of the
form
 \be
ds^2 \sim -\frac{f(x)}{\Sigma(x)}\,dt +
\frac{\kappa^2\Sigma(x)(1-x^2)}{f(x)} \left[\frac{dx^2}{1-x^2} +
\frac{dy^2}{y^2-1} + (y^2-1)d\varphi^2\right],
 \ee
with $f(x)$ strictly positive. Axial ($l=0$) geodesics along
$y=\pm1$ connect the two horizons in region $II$, the first
integrated geodesic equation being (\ref{geo2}). For timelike axial
geodesics, it seems (see Appendix C) that the effective potential
$V(x)$ has a relative maximum in the range $x\in[-1,0]$ with
$V_{max} > V(\infty) = 1/\kappa^2$. So an observer radially
infalling from $x=+\infty$ with sufficiently high velocity will
cross the two horizons and proceed towards $x=-\infty$ in region
$III$, but an observer with sufficiently low velocity will instead
be reflected back through the outer horizon to $x = + \infty$,
albeit in another spacetime coordinate patch $I$, to the future of
the previous one. The effective radial distance between the two
horizons in region $II$ is $\Delta r = \kappa\Delta x = 2\kappa =
pM$.

When $|y|$ increases, an ergosurface $f(x,y)=0$ appears. The
behaviors of the various metric and electromagnetic functions for
$y^2\to\infty$
 \ba\lb{yinfty}
f &\sim& -\frac{q^2x^2y^2}{1-x^2}, \quad \Sigma \sim \v \sim
\frac{q^4y^4}{4(1-x^2)^2}, \quad -e^{2\nu} \sim
\frac{4x^2(1-x^2)^2}{p^2y^6}, \nn\\ \Pi &\sim& \Pi_2(x)y^4, \quad
\Theta \sim \frac{q^5y^6}{8x(1-x^2)^2}
 \ea
lead to the non-asymptotically flat behavior of the metric and
electromagnetic field
 \ba
ds^2 &\sim& -\frac{\kappa^2q^2y^4}{4x^2}d\varphi^2 +
\frac{\kappa^2q^4x^2}{p^2y^2}\left(\frac{dy^2}{y^2} +
\frac{dx^2}{1-x^2} + \right. \nn\\
&+& \left. \frac{4p^2(1-x^2)}{\kappa^2q^6}\left[dt +
\frac{\kappa\Pi_2(x)(1-x^2)y^2}{q^2x^2}d\varphi\right]^2 \right),\lb{S0met}\\
A &\sim&
\varepsilon\left\{\left[1+\frac{(1-x^2)(p-6x-3px^2+4x^3)}{q^2xy^2}\right]dt
+ \frac{\kappa qy^2}{2x}\,d\varphi\right\}. \lb{S0em}
 \ea
The squared Ricci scalar diverges as $y^4$.

As it is enclosed between two horizons which are topological
spheres, the singularity $|y|\to\infty$ must actually be at finite
distance. Indeed, putting $y=\zeta^{-1}$ and $x=\cos\chi$, the
asymptotic metric (\ref{S0met}) takes the form
 \ba
ds^2 &\simeq& -\frac{\kappa^2q^2}{4\cos^2\chi}\zeta^{-4}\,d\varphi^2
+ \frac{\kappa^2q^4\cos^2\chi}{p^2}\left[d\zeta^2 + \zeta^2\,d\chi^2 \right. \nn\\
&+& \left. \frac{4p^2\zeta^2}{\kappa^2q^6}\sin^2\chi(dt +
\frac{\kappa\Pi_2(\chi)\sin^2\chi}\zeta^{-2}d\varphi^2)\right],
 \ea
It follows that $\zeta=0$ corresponds to a point $S_0$ in the
$\varphi=$ constant sections, or a closed timelike line of the
four-dimensional spacetime. Note also that $\sqrt{|g|}$ goes to the
finite limit $(\kappa^3q^4/p^2)\cos^2\chi|\sin\chi|$ for $\zeta \to
0$, so that this singularity has a finite volume per unit time
$(4\pi/3)\kappa^3q^4/p^2$.

\begin{figure}[tb]
\begin{center}
\begin{minipage}[t]{0.58\linewidth}
\hbox to\linewidth{\hss%
  \includegraphics[width=1.5\linewidth,height=1.2\linewidth]{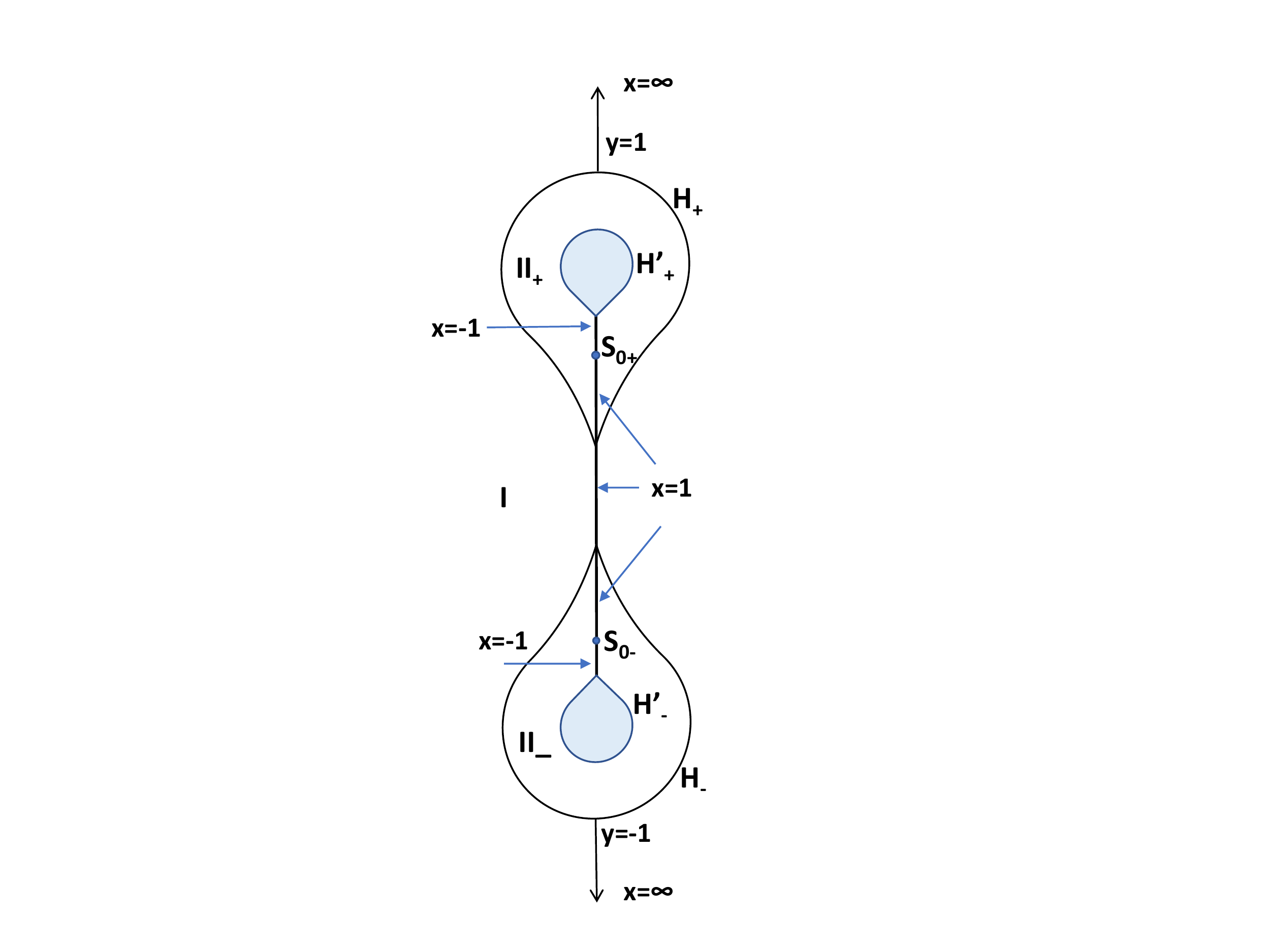}
\hss}
\end{minipage}
 \caption{\small Relative positions of the two outer and inner horizons,
the connecting strings in regions $I$ and $II_\pm$, and the singularities
$S_{0\pm}$.} \label{F8}
\end{center}
\end{figure}

For $E\neq0$, the effective potential $U$ is dominated by the term
$-E^2/F$, which is positive and increases as $y^2$, so that the
geodesics turn back before reaching $y\to\infty$. For $E=0$,
however, $U=L^2F/\rho^2$ is negative and goes to zero as $y^{-4}$,
so that timelike geodesics again turn back while null geodesics
extend to infinity and are complete ($\tau \propto y$). Thus only
spacelike geodesics with $E=0$ terminate at the singularity $S_0$.
This analysis applies equally to geodesics following the cosmic
strings $x=1$ and $x=-1$. Indeed, analytic extension of the
geodesic motion through the two exterior horizons $H_+$ and $H_-$
shows that the line $x=1$ may be viewed as a single cosmic string
connecting the two singularities $S_{0\pm}$. Likewise, the line
$x=-1$ may be considered as a single cosmic string connecting the
two singularities $S_{0\pm}$ through the two interior horizons
$H'_+$ and $H'_-$, these singularities themselves arising from the
mismatch between the different tensions and angular velocities of
the two cosmic strings. These cosmic strings are also Dirac and
Misner strings, the exterior cosmic string carrying the magnetic and
gravimagnetic fluxes computed in Sect. 3, and the
interior cosmic string carrying the corresponding fluxes with $p$
replaced by $-p$. It follows that the singularities $S_{0\pm}$ have
opposite magnetic charges $\pm P_0$ and NUT charges $\pm N_0$ with
 \be
P_0 = \frac{\varepsilon\kappa}{2q}[\gamma(p)-\gamma(-p)] =
\frac{\varepsilon\kappa}{q}(3+p^2), \quad  N_0 =
\frac{\kappa}{4q}[\lambda(p)-\lambda(-p)] = \frac{\kappa}{q}(1+p^2).
 \ee
 %%%%%%%%%%%%%%%%%%%%%Fig8
Fig. \ref{F8} shows the relative positions of the two outer and inner horizons,
the connecting strings in regions $I$ and $II_\pm$, and the singularities
$S_{0\pm}$.
%%%%%%%%%%%%%%%%%%%%%%%%%%%%%
The properties of the third region $III$ ($x<-1$ and $-1\le y\le1$)
are in the whole similar to those of the exterior region $I$, except
for the existence of the ring singularity $\Sigma(x,y)=0$, i.e.
$y=0$ and $x=x_0$ with
 \be\lb{Sig0}
\frac{p(x_0^2+1)}{2x_0} + 1 + \frac{q^2}{2(x_0^2-1)} = 0.
 \ee
We show in Appendix C that, for every $p \in ]0,1[$, this equation
has a solution $x_0(p) < -1$. This is actually such that $x_0(p) <
-1/p$ . To the difference of the case of Kerr, $f(x_0,0)$ does not
vanish. Using (\ref{Sig0}), one can show that
 \be
f(x_0,0) = \frac{(px_0^2+2x_0+p)^2}{4x^2} > 0,
 \ee
so that the equatorial ring singularity is timelike. From (\ref{U})
the effective potential $U$ diverges on the ring, so that all
geodesics are generically repelled away from the ring by a potential
barrier. The only geodesics which can reach the ring are spacelike
geodesics with parameters fine-tuned so that $L/E =
\kappa\Pi(x_0)/f(x_0)$, in which case $L-E\omega$ goes to zero as
$x-x_0$, so that $U$ goes on the ring to a constant positive value
$U_0$. Note also that on this ring $g_{\varphi\varphi} \sim
-\kappa^2\Pi^2/\Sigma f \to -\infty$. As a consequence,
$g_{\varphi\varphi}$ must be negative in the vicinity of the ring,
which therefore lies within the region of CTCs.

\begin{figure}[tb]
\begin{center}
\begin{minipage}[t]{0.68\linewidth}
\hbox to\linewidth{\hss%
  \includegraphics[width=1.5\linewidth,height=1.2\linewidth]{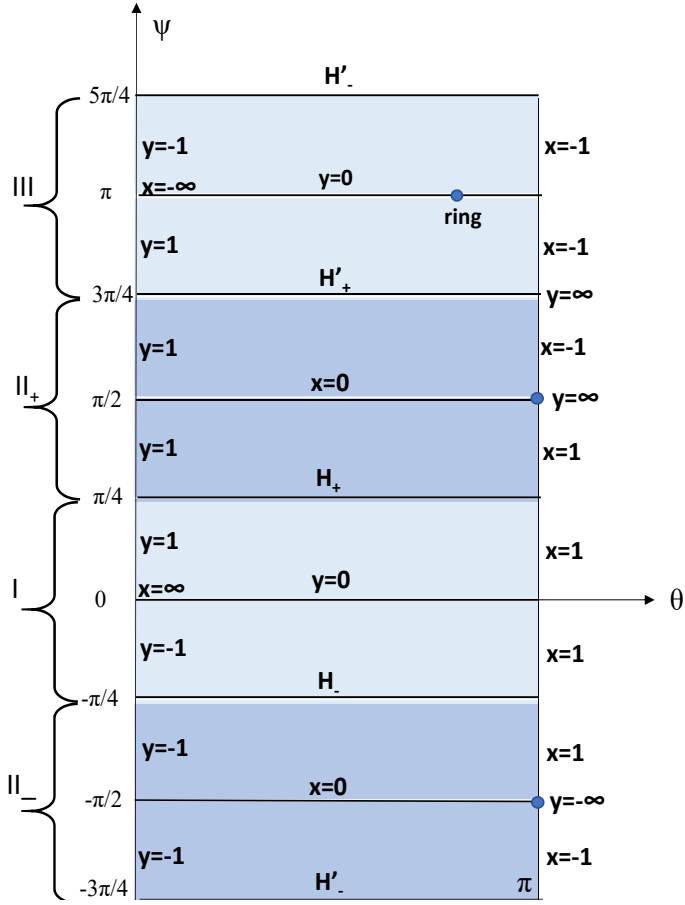}
\hss}
\end{minipage}
\caption{\small Spatial topology of the two-black hole system, with the
outer regions $I$ and $III$ and inner regions $II_\pm$. The horizons
$H_\pm$ and $H'_\pm$ are represented by horizontal lines, the two
regular $z$ axes by the left vertical line, and the two singular
strings by the right vertical line.} \label{F9}
\end{center}
\end{figure}

If the two innermost regions $III$ are identified, the spatial
topology of the maximally extended spacetime can be described, in
terms of the coordinates $X$ (or $\eta=2\arctan{X}$) and $Y$
previously introduced, as that of a truncated cylinder, with
longitudinal coordinate $X \ge 0$ ($0\le\eta\le\pi$) and angular
coordinate $\psi$ related to $Y$ by $Y=\tan\psi$, see Fig. \ref{F9}.
%%%%%%%%%%%%%%%%%%%%%%%Fig9F
%%%%%%%%%%%%%%%%%%%%%%%%%%
The basis circle
$X=0$ ($\eta=0$) corresponds to the regular portion of the
symmetry axis, $y^2=1$, except for the two points $\psi=0$
($x\to+\infty$) and $\psi=\pi$ ($x\to-\infty$) which correspond to
spacelike infinity. So the regular portion of the symmetry axis
has two components, each homeomorphic to the real line.
The generatrices $\psi=0$ or $\psi=\pi$
correspond to the equatorial plane. The horizons are represented by
the generatrices $\psi=\pi/4 + k\pi/2$, the sector
$\psi\in]-\pi/4,\pi/4[$ corresponding to region $I$, the sector
$\psi\in]\pi/4,3\pi/4[$ to region $II_+$, the sector
$\psi\in]3\pi/4,5\pi/4[$ to region $III$, and the sector
$\psi\in]5\pi/4,7\pi/4[$ to region $II_-$. The circle $X\to\infty$
($\eta=\pi$) corresponds to the singular cosmic strings $x^2=1$
connecting together, through the two exterior or interior horizon
conical singularities, the two singular points $S_{0\pm}$
($X\to\infty$, $\psi=\pm\pi/2$) which correspond to the timelike
singularities $y\to\pm\infty$ in regions $II_\pm$. Finally, the
isolated timelike singular ring $(x=x_0,y=0)$ is represented by a
point $X=X_0$ on the generatrix $\psi=\pi$.

The metric function $f=-||\partial/\partial t||^2$ is positive on
the basis circle $X=0$, while it is negative on the horizons
$\psi=\pi/4 + k\pi/2$. So the $\varphi=$ constant sections of the
ergosurfaces are represented by curves connecting the successive
points ($X=0$, $\psi=\pi/4 + k\pi/2$), the ergosphere extending from
these curves to $X\to\infty$. Conversely,
$g_{\varphi\varphi}=||\partial/\partial\varphi||^2$ is positive on
the horizons and negative on the singular circle $X\to\infty$, so
that the causal boundary $g_{\varphi\varphi}=0$ is represented by
curves connecting successive points ($X=\to\infty$, $\psi=\pi/4 +
k\pi/2$), the domain extending from these curves to $X\to\infty$
containing CTCs. Because the singular ring ($X=X_0$, $\psi=\pi$)
belongs both to the stationary domain $f>0$ and to the domain
containing CTCs, the ergosurface and the causal boundary must
intersect in region $III$.

\section{Conclusions}
The  binary black hole presented here differs from numerous previously known
double-center solutions in many respects. First, it was derived without use of ISM,
which is designed to generate solutions with as many independent parameters as possible.
This generality, however, creates problems with calculating physical parameters and
revealing physical properties of the solutions.
Our solution is much more simple and contains only two parameters,
which can be chosen as the total mass and the total angular momentum, exactly as in the Kerr case.
It was derived using an original generating technique due to one of the authors which is a product of
invariance transformations of the dimensionally reduced target space sigma model with a rotation in
the space of Killing orbits. Applied to the ZV2 vacuum metric, this procedure leads to a rotating
two-center solution of the Einstein-Maxwell equations with many attractive features.
Asymptotically it looks like the Kerr solution with zero Coulomb charges but with magnetic dipole
and electric quadrupole moments. It has an ergosphere inside which one finds two extremal
co-rotating black holes touching the ergosphere at the points on the symmetry axis,
endowed with equal electric charges and opposite magnetic and NUT charges.
Since the total NUT charge is zero, the metric is asymptotically flat.

Contrary to many known rotating double-center solutions to vacuum and electrovacuum gravity,
our solution is manifestly free from ring singularities outside the horizons. Still, the solution is
unbalanced and contains conical singularities on the segment of the polar axis between
the constituent black holes. This string co-rotates with the horizons, has an electric charge balancing the two
black hole charges, and is also the Dirac string carrying the magnetic flux between the opposite
magnetic monopoles, and the Misner string carrying the gravimagnetic flux between the
opposite NUT charges.

The rotating string is surrounded by a tiny chronosphere which lies entirely inside the ergosphere,
its maximal size being of the order of $10^{-4}$ of the length  defined by the Schwarzschild mass
of the solution. We investigated its structure finding that one of the boundaries of the dragging
angular velocity diverges on the chronosphere. Both boundaries of dragging  velocities converge
on the polar axis to a zero value.

The solution was analytically continued inside the horizons. The most economical
maximal analytical extension contains two isometrical interior regions between
an outer and an inner horizon (both degenerate). Inside these interior regions one meets a strong closed
timelike singularity. Beyond the inner horizons there is a third, asymptotically flat region
containing a timelike ring singularity repelling almost all geodesics.

Our family of solutions interpolates between the vacuum ZV2 solution and (after a suitable rescaling)
the extreme Kerr metric. Remarkably, the total horizon area of the constituents is, in this extreme limit,
one-half of the horizon area of the limiting Kerr black hole of the same mass. This is similar to the
case of ZV2: as shown by Kodama and Hikida \cite{Kodama:2003ch}, the total horizon area of the
two constituents is one-half of the area of the Schwarzschild black hole of the same mass.
We leave a more complete  discussion of thermodynamics for future work.

\section*{Acknowledgments} DG thanks LAPTh Annecy-le-Vieux
for hospitality at different stages of
this work. He also acknowledges the support of the Russian
Foundation of Fundamental Research under the project 17-02-01299a
and the Russian Government Program of Competitive Growth of the
Kazan Federal University.

\renewcommand{\theequation}{A.\arabic{equation}}
\setcounter{equation}{0}
\section*{Appendix A: Positivity of $g_{\varphi\varphi}$ on the ergosurface}
Let us show that
 \be
g_{\varphi\varphi} = \kappa^2\left(\frac{\Sigma}f(x^2-1)(1-y^2) -
\frac{\Pi^2}{\Sigma f}\right)
 \ee
is finite on the ergosurface where
 \be\lb{fA}
f(x,y) =  \frac{p^2(x^2-1)^2}{4x^2} - \frac{q^2x^2(1-y^2)}{x^2-1}
 \ee
vanishes.

Putting $x^2-1=\xi^2$, $1-y^2=\eta^2$, $g_{\varphi\varphi}$ factors as
 \be\lb{gpp1}
g_{\varphi\varphi} = \frac{\kappa^2\xi^2\eta^2}{\Sigma
f}\left(\Sigma-\frac{\Pi}{\xi\eta}\right)\left(\Sigma+\frac{\Pi}{\xi\eta}\right)
= \frac{\kappa^2\xi^2\eta^2}{\Sigma f}\,\Sigma_-\,\Sigma_+\,.
 \ee
From (\ref{Pi}), we may expand
 \be
\frac{\Pi}{\xi\eta} = \frac{\Pa(x)}\xi\,\beta + \frac{\Pb(x)}\xi\,\beta^3 \qquad (\beta \equiv q\eta),
 \ee
where we have put $\Pi_1 = q\Pa$, $\Pi_2 = q^3\Pb$. Similarly, $\Sigma$
given by (\ref{Sig}) may be expanded as
 \be
\Sigma = \Sigma_0(x) + \Sigma_2(x)\beta^2 + \Sigma_4(x)\beta^4.
 \ee
Assuming without loss of generality $q>0$, Eq. (\ref{fA}) may be
inverted and linearized near $f=0$ to:
 \be\lb{EA}
\beta =  \beta_0(x) - \frac{f}{p\xi} + {\rm O}(f^2), \quad \beta_0 \equiv
\frac{p\xi^3}{2x^2}.
 \ee
We can then show that the function
 \be
\Sigma_+(x,\beta) \equiv \Sigma_0(x) + \frac{\Pa(x)}\xi\beta +
\Sigma_2(x)\beta^2 + \frac{\Pb(x)}\xi\beta^3 + \Sigma_4(x)\beta^4
 \ee
vanishes identically for $\beta =  \beta_0(x)$
($f=0$), so that $g_{\varphi\varphi}$ is given on the ergosurface by
 \be\lb{gE}
\left.g_{\varphi\varphi}\right\vert_E = - \frac{\kappa^2p\xi^7}{2q^2x^4}\,\frac{\partial\Sigma_+}
{\partial\beta}\left(x,\beta_0(x)\right).
 \ee

Evaluation of (\ref{gE}) leads to
 \ba
\left.g_{\varphi\varphi}\right\vert_E &=& \frac{\kappa^2p\xi^4}{2q^2x^{10}}\left\{
\frac{x^3}{4}\left[8x^4(x^2+p^2) - p^2\xi^6\right]\right. \nn\\
&& \left. + \left[\frac{8+p^2-p^4}{2p}x^6 + \frac4p\xi^2x^6 +
\frac{p}{32}\xi^6(32x^4-5p^2x^2+p^2)\right]\right\}.
 \ea
Both functions of $x$ inside square brackets are positive definite for $x^2>1$, $0<p<1$,
so that $\partial_\varphi$ is spacelike on the ergosurface in the outer region $I$ ($x>1$).
On the other hand, in the innermost region $III$ ($x<-1$), either term
may dominate depending on the value of $x$, meaning that the ergosurface
and causal boundary may intersect.

\renewcommand{\theequation}{B.\arabic{equation}}
\setcounter{equation}{0}
\section*{Appendix B: Relative maximum of the effective potential $V(x)$ in region $II$}
The effective potential $V(x) = p^2(x^2-1)^2/4\kappa^2x^2\Sigma(x)$
for timelike axial geodesics in the interior region II
($-1<x<1$, $\epsilon=-1$ in (\ref{geo2})) is
larger than the potential at infinity $V_{\infty} = 1/\kappa^2$ if
$F(x)<0$, where $F(x)$ is the cubic
 \be
F(x) \equiv 4px^3 + 8x^2 + 4p^3x + p^2q^2.
 \ee
$F$ is positive for $x=0$ and $x=-1$ ($F(-1)=4(1-p)(p^2+p+2)$),
while its derivative
 \be
F'(x) = 4(3px^2+4x+p^3)
 \ee
is positive for $x=0$ and negative for $x=-1$
($F'(-1)=-4(1-p)(p^2+p+4)$). So $F$ must have a minimum $x_0$
somewhere in the range $x\in[-1,0]$. If this minimum value is
negative, then $V(x_0)>V_{\infty}$.

It is not clear that $F(x_0)<0$ for all $p\in[0,1]$. But it is easy
to show that for $|q|$ small enough, one can find some $x\in[-1,0]$
such that $F(x)<0$. Take e.g. $x=-p/2$. Then,
 \be
F(-p/2) = 3p^2(q^2-p^2/6)
 \ee
is negative for $q^2<1/7$.

\renewcommand{\theequation}{C.\arabic{equation}}
\setcounter{equation}{0}
\section*{Appendix C: Existence of the ring singularity in region $III$}
In the equatorial plane, $\Sigma(x,0) = F^2(x)/4x^2(x^2-1)^2$ with
 \be
F(x) \equiv px^4 + 2x^3 - (1+p^2)x - p.
 \ee
Its derivative $F'(x)$ is negative for $x\to-\infty$, has a local
maximum at $x=-1/p$, where it is positive, and a local minimum at
$x=0$. It is also positive for $x=-1$. It must be therefore negative
for $x<\alpha$ and remain positive in the range $\alpha <x \le -1$,
for some $\alpha<-1/p$. So $F(x)$ decreases from $x\to-\infty$,
where it is positive, to $X=\alpha$, then increases to $x=-1$, where
it is negative. It follows that $F(x)$ must vanish once in the range
$x\le-1$ for some value $x=x_0<\alpha<-1/p$.


\begin{thebibliography}{9}

 %\cite{Barack:2018yly}
\bibitem{Barack:2018yly}
  L.~Barack {\it et al.},
  %``Black holes, gravitational waves and fundamental physics: a roadmap,''
  arXiv:1806.05195 [gr-qc].

%\cite{Yagi}
\bb{Yagi}
 K.~Yagi and L.~C.~Stein,
  %``Black Hole Based Tests of General Relativity,''
 Class.\ Quant.\ Grav.\  {\bf 33}, 054001 (2016)
 [arXiv:1602.02413 [gr-qc]].

%\cite{Cunha:2018acu}
\bibitem{Cunha:2018acu}
  P.~V.~P.~Cunha and C.~A.~R.~Herdeiro,
  %``Shadows and strong gravitational lensing: a brief review,''
  Gen.\ Rel.\ Grav.\  {\bf 50}, no. 4, 42 (2018).
  doi:10.1007/s10714-018-2361-9
  [arXiv:1801.00860 [gr-qc]].

 %\cite{Kramer1980}
\bibitem{Kramer1980}
 D.~Kramer and G.~Neugebauer,
%`` The superposition of two Kerr solutions, ``
 Phys.\ Left.\ {\bf A 75} (1980), 259.

%\cite{Costa:2010zzg}
\bibitem{Costa:2010zzg}
  M.~S.~Costa, C.~A.~R.~Herdeiro and C.~Rebelo,
  %``Physical properties of the double Kerr solution,''
  J.\ Phys.\ Conf.\ Ser.\  {\bf 229}, 012062 (2010).
  doi:10.1088/1742-6596/229/1/012062.

 %\cite{Cunha:2018gql}
\bibitem{Cunha:2018gql}
  P.~V.~P.~Cunha, C.~A.~R.~Herdeiro and M.~J.~Rodriguez,
  %``Does the black hole shadow probe the event horizon geometry?,''
  Phys.\ Rev.\ D {\bf 97}, no. 8, 084020 (2018).
  doi:10.1103/PhysRevD.97.084020
  [arXiv:1802.02675 [gr-qc]].

%\cite{Cunha:2018cof}
\bibitem{Cunha:2018cof}
  P.~V.~P.~Cunha, C.~A.~R.~Herdeiro and M.~J.~Rodriguez,
  %``Shadows of Exact Binary Black Holes,''
  arXiv:1805.03798 [gr-qc].

  %\cite{Johannsen:2013rqa}
\bibitem{Johannsen:2013rqa}
  T.~Johannsen,
  %``Systematic Study of Event Horizons and Pathologies of Parametrically Deformed Kerr Spacetimes,''
  Phys.\ Rev.\ D {\bf 87}, no. 12, 124017 (2013).
  doi:10.1103/PhysRevD.87.124017
  [arXiv:1304.7786 [gr-qc]].

  %\cite{Clement:2017kas}
\bibitem{Clement:2017kas}
  G.~Cl{\'e}ment and D.~Gal'tsov,
  %``A tale of two dyons,''
  Phys.\ Lett.\ B {\bf 771}, 457 (2017).
  doi:10.1016/j.physletb.2017.05.096
  [arXiv:1705.08017 [gr-qc]].

\bb{GC98} G. Cl{\'e}ment, Phys.\ Rev.\ D {\bf 37}, 4885 (1998)
[arXiv:gr-qc/9710109].

%\cite{Bonnor:1966}
\bibitem{Bonnor:1966}  W.~B.~Bonnor,
%``An exact  solution of the Einstein-Maxwell equations referring to a magnetic dipole,''
Z. Phys. {\bf 190}, 444 (1966).

 %\cite{Emparan:1999au}
\bibitem{Emparan:1999au}
  R.~Emparan,
  %``Black diholes,''
  Phys.\ Rev.\ D {\bf 61}, 104009 (2000).
  doi:10.1103/PhysRevD.61.104009
  [hep-th/9906160].

 %\cite{Emparan:2001bb}
\bibitem{Emparan:2001bb} R.~Emparan and E.~Teo, %``Macroscopic and microscopic description of black diholes,''
Nucl.\ Phys.\ B {\bf 610}, 190 (2001).
doi:10.1016/S0550-3213(01)00319-4
[hep-th/0104206].

%\cite{Tomimatsu:1972zz}
 \bibitem{Tomimatsu:1972zz} A.~Tomimatsu and H.~Sato,
 %``New Exact Solution for the Gravitational Field of a Spinning Mass,''
Phys.\ Rev.\ Lett.\ {\bf 29}, 1344 (1972).
doi:10.1103/PhysRevLett.29.1344;
  Progr. Theor. Phys. {\bf 50}, 95 (1973).

\bb{BW}
 R. Bach and H. Weyl, Math. Z. {\bf 13}, 134 (1922).

\bb{darmois} G. Darmois,
 %`` Les \'equations de la gravitation einsteinienne'',
M\'emorial des sciences math\'ematiques, Fasc. XXV,
Gauthiers-Villars, Paris 1927.

%\cite{Zipoy1966,Voorhees:1971wh}
\bibitem{Zipoy1966} D.M. Zipoy,
%``Topology of some spheroidal matter'',
J.\ Math.\ Phys.  {\bf 7}, 1137 (1966).
doi.org/10.1063/1.1705005.

%\cite{Voorhees:1971wh}
\bibitem{Voorhees:1971wh} B.~H.~Voorhees,
 %``Static axially symmetric gravitational fields,''
Phys.\ Rev.\ D {\bf 2}, 2119 (1970).
doi:10.1103/PhysRevD.2.2119.

\bb{gibbons73} G.W. Gibbons and R. Russell-Clark, Phys. Rev. Lett.
{\bf 30}, 398 (1973).

%\cite{Ernst}
\bibitem{Ernst}
F.J. Ernst,
%``New representation of the Tomimatsu-Sato solution `'
J. Math Phys. {\bf 17}, 1091 (1976).

%\cite{Economou}
\bibitem{Economou}
J. E. Economou,
%``Approximate form of the Tomimatsu-Sato 0 = 2 solution near the poles x=1, y= ±1 `',
J. Math Phys. {\bf 17}, 1095 (1976).

%\cite{Papadopoulos:1981wr}
\bibitem{Papadopoulos:1981wr}
  D.~Papadopoulos, B.~Stewart and L.~Witten,
  %``Some Properties of a Particular Static, Axially Symmetric Space-time,''
  Phys.\ Rev.\ D {\bf 24}, 320 (1981).
  doi:10.1103/PhysRevD.24.320.

%%\cite{Manko:1999xg}
\bibitem{Manko:1999xg}
  O.~V.~Manko, V.~S.~Manko and J.~D.~Sanabria-G{\'o}mez,
  %``Remarks on the charged, magnetized Tomimatsu-Sato delta = 2 solution,''
  Gen.\ Rel.\ Grav.\  {\bf 31}, 1539 (1999).
  doi:10.1023/A:1026782404418.

  %\cite{Kodama:2003ch}
\bibitem{Kodama:2003ch} H.~Kodama and W.~Hikida, %``Global structure of the Zipoy-Voorhees-Weyl spacetime and the delta=2 Tomimatsu-Sato spacetime,''
Class.\ Quant.\ Grav.\ {\bf 20}, 5121 (2003).
doi:10.1088/0264-9381/20/23/011
[gr-qc/0304064].

%\cite{Gegenberg:2010np}
\bibitem{Gegenberg:2010np} J.~Gegenberg, H.~Liu, S.~S.~Seahra and B.~K.~Tippett, %``Tomimatsu-Sato geometries, holography and quantum gravity,''
Class.\ Quant.\ Grav.\ {\bf 28}, 085004 (2011).
doi:10.1088/0264-9381/28/8/085004
[arXiv:1010.2803 [hep-th]].

\bibitem{Ernst1}
F.J. Ernst,
%``Charged version of Tomimatsu-Sato spinning mass field `'
 Phys. Rev. {\bf D7}, 2510 (1973).

%\cite{Hoenselaers:1985qk}
\bibitem{Hoenselaers:1985qk}
 D. Kramer and G. Neugebauer, in
  ``Solutions of Einstein's Equations: Techniques and Results. Proceedings, International Seminar, Retzbach,
  F.R. Germany, November 14-18, 1983,'' W.~Dietz and  C.~Hoenselaers, eds.,
Lecture Notes in Physics, v.205, 1984.

   %\cite{Belinsky:1979mh,Belinski:2001ph}
\bibitem{Belinsky:1979mh}
  V.~A.~Belinsky and V.~E.~Zakharov,
  %``Stationary Gravitational Solitons with Axial Symmetry,''
  Sov.\ Phys.\ JETP {\bf 50}, 1 (1979)
  [Zh.\ Eksp.\ Teor.\ Fiz.\  {\bf 77}, 3 (1979)].

%\cite{Belinski:2001ph}
\bibitem{Belinski:2001ph}
  V.~Belinski and E.~Verdaguer,
  ``Gravitational Solitons,''
  CUP, United Kingdom (2005).
ISBN 10: 0521018064, ISBN 13: 9780521018067,
  doi:10.1017/CBO9780511535253.

%\cite{Herdeiro:2008kq}
\bibitem{Herdeiro:2008kq}
  C.~A.~R.~Herdeiro and C.~Rebelo,
  %``On the interaction between two Kerr black holes,''
  JHEP {\bf 0810}, 017 (2008).
  doi:10.1088/1126-6708/2008/10/017
  [arXiv:0808.3941 [gr-qc]].

\bb{GP} J.B. Griffiths and J. Podolsky, ``Exact Space-times in Einstein's General Relativity'', CUP, 2009.

%\cite{Alekseev:2017zuh}
\bibitem{Alekseev:2017zuh}
  G.~A.~Alekseev,
  %``Integrable and non-integrable structures in Einstein - Maxwell equations with Abelian isometry group $\mathcal{G}_2$,''
  Trudy Steklov Mat.\ Inst.\  {\bf 295}, 7 (2016)
  doi:10.1134/S0081543816080010
  [arXiv:1702.05925 [gr-qc]].

   %\cite{Breitenlohner:1987dg}
\bibitem{Breitenlohner:1987dg}
  P.~Breitenlohner, D.~Maison and G.~W.~Gibbons,
  %``Four-Dimensional Black Holes from Kaluza-Klein Theories,''
  Commun.\ Math.\ Phys.\  {\bf 120}, 295 (1988).
  doi:10.1007/BF01217967.

  %\cite{Bonnor:1993,Perry}
\bibitem{Bonnor:1993}  W.~B.~Bonnor,
%``The equilibrium of a charged test particle in the field of a spherical charged mass in general relativity,''
Z. Phys. {\bf 190}, 444 (1966).
 Class. Quant. Grav. {\bf 10}, 2077 (1993).

 %\cite{Perry}
\bb{Perry}
  G. P. Perry and F. I. Cooperstock
  %”Electrostatic equilibrium of two spherical charged masses in general relativity”,
  Class. Quant. Grav. {\bf 14}, 1329 (1997); arXiv:gr-qc/9611066.

 %\cite{Tomimatsu:1981bc}
\bibitem{Tomimatsu:1981bc}
  A.~Tomimatsu and M.~Kihara,
  %``Conditions for Regularity on the Symmetry Axis in a Superposition of Two Kerr Nut Solutions,''
  Prog.\ Theor.\ Phys.\  {\bf 67}, 1406 (1982).
  doi:10.1143/PTP.67.1406.

  %\cite{Kihara:1981mx}
\bibitem{Kihara:1981mx}
  M.~Kihara and A.~Tomimatsu,
  %``Some Properties of the Symmetry Axis in a Superposition of Two Kerr Solutions,''
  Prog.\ Theor.\ Phys.\  {\bf 67} (1982) 349.
  doi:10.1143/PTP.67.349.

%\cite{Tomimatsu:1983qc}
 \bibitem{Tomimatsu:1983qc} A.~Tomimatsu,
  %``On Gravitational Mass and Angular Momentum of Two Black Holes in Equilibrium,''
Prog.\ Theor.\ Phys.\ {\bf 70}, 385 (1983).
doi:10.1143/PTP.70.385

%\cite{Dietz1985}
\bibitem{Dietz1985}
  C.~Hoenselaers and W.~Dietz,
  %``Two Mass Solutions of Einstein’s Vacuum Equations: The Double Kerr Solution,''
 Annals of Physics, {\bf 165}, 319-383 (1985).

\bibitem{Neugebauer:2009su}
  G.~Neugebauer and J.~Hennig,
  %``Non-existence of stationary two-black-hole configurations,''
  Gen.\ Rel.\ Grav.\  {\bf 41}, 2113 (2009).
  doi:10.1007/s10714-009-0840-8
  [arXiv:0905.4179 [gr-qc]].

%\cite{Hennig:2011fp}
\bibitem{Hennig:2011fp}
  J.~Hennig and G.~Neugebauer,
  %``Non-existence of stationary two-black-hole configurations: The degenerate case,''
  Gen.\ Rel.\ Grav.\  {\bf 43}, 3139 (2011).
  doi:10.1007/s10714-011-1228-0
  [arXiv:1103.5248 [gr-qc]].

%\cite{Neugebauer:2011qb}
\bibitem{Neugebauer:2011qb}
  G.~Neugebauer and J.~Hennig,
  %``Stationary two-black-hole configurations: A non-existence proof,''
  J.\ Geom.\ Phys.\  {\bf 62}, 613 (2012).
  doi:10.1016/j.geomphys.2011.05.008
  [arXiv:1105.5830 [gr-qc]].

 %\cite{Chrusciel:2011iv}
\bibitem{Chrusciel:2011iv}
  P.~T.~Chrusciel, M.~Eckstein, L.~Nguyen and S.~J.~Szybka,
  %``Existence of singularities in two-Kerr black holes,''
  Class.\ Quant.\ Grav.\  {\bf 28}, 245017 (2011).
  doi:10.1088/0264-9381/28/24/245017
  [arXiv:1111.1448 [gr-qc]].

%\cite{Alekseev:2012au}
\bibitem{Alekseev:2012au}
  G.~A.~Alekseev and V.~A.~Belinski,
  %``On the equilibrium state of two rotating charged masses in General Relativity,''
  arXiv:1211.3964 [gr-qc].

%\cite{Kinnersley:1978pz,Kinnersley:1978}
\bibitem{Kinnersley:1978pz}
  W.~Kinnersley and D.~M.~Chitre,
  %``Symmetries of the Stationary Einstein-Maxwell Field Equations. III.,''
  J.\ Math.\ Phys.\  {\bf 19}, 1926 (1978).
  doi:10.1063/1.523912

%\cite{Kinnersley:1978}
\bibitem{Kinnersley:1978}
  W.~Kinnersley and D.~M.~Chitre,
  %``Symmetries of the Stationary Einstein-Maxwell Field Equations. IV,''
  J.\ Math.\ Phys.\  {\bf 19}, 2037 (1978).

\bibitem{manko00} V.S.~Manko, E.W. Mielke and J.D.~Sanabria-G{\'o}mez,
Phys.\ Rev.\ D {\bf 61} (2000). 081501 [arXiv:gr-qc/0001081].

\bb{BG} J.D. Barrow and G.W. Gibbons,
%``Maximum magnetic moment to angular momentum conjecture,''
Phys.\ Rev.\ D {\bf 95}, no. 6, 064040 (2017).
doi:10.1103/PhysRevD.95.064040,
[arXiv:1701.06343].

\bibitem{RTN}
  G.~Cl\'ement, D.~Gal'tsov and M.~Guenouche,
  %``Rehabilitating space-times with NUTs,''
 Phys.\ Lett.\ B {\bf 750}, 591 (2015) [arXiv:1508.07622 [hep-th]].

\bibitem{NW}
  G.~Cl\'ement, D.~Gal'tsov and M.~Guenouche,
  %``NUT wormholes,''
 Phys.\ Rev.\ D {\bf 93}, 024048 (2016) [arXiv:1509.07854[hep-th]].

\bb{DJH} S. Deser, R. Jackiw and G. 't Hooft, Annals Phys. {\bf
152}, 220 (1984).

\bb{GC85} G. Cl\'ement, Int. J. Theor. Phys {\bf 24}, 267 (1985).

\bb{tom84} A. Tomimatsu, Progr. Theor. Phys. {\bf 72}, 73 (1984).

\bb{smarr}
 G. Cl\'ement and D. Gal'tsov, Phys. Lett. B {\bf 773} (2017) 290
[arXiv:1707.01332].

\end{thebibliography}
\end{document}